\documentclass[twocolumn]{aastex63}

\usepackage{times}

\usepackage{graphicx}

\usepackage{tikz}

\usepackage[T1]{fontenc}
\usepackage{aecompl}
\usepackage{calligra}
\DeclareMathAlphabet{\mathcalligra}{T1}{calligra}{m}{n}
\DeclareFontShape{T1}{calligra}{m}{n}{<->s*[2.2]callig15}{}

\usepackage{hyperref}
\hypersetup{
    pdfnewwindow=true,      
    colorlinks=true,       
    linkcolor=orange,      
    citecolor=cyan,        
    filecolor=orange,      
    urlcolor=orange           
}

\defcitealias{DL:2016}{DL16}

\usepackage{color}

\def\orb{\rm{orb}}
\def\obs{\rm{obs}}
\def\min{\rm{min}}
\def\max{\rm{max}}

\def\Msun{{M_{\odot}}}

\def\MBH{M_{\rm{BH}}}
\def\MNS{M_{\rm{NS}}}
\def\Lum{\mathcal{P}}

\def\Rbin{r}
\def\NS{\mathrm{NS}}
\def\BH{\mathrm{BH}}
\def\RNS{R_{\NS}}
\def\Res{\mathcal{R}}

\def\eg{{\em e.g.}}

\begin{document}

\shorttitle{Multi-Messenger Magnetic Field Constraints}
\shortauthors{D'Orazio, Haiman, Levin, Samsing, $\&$ Vigna-G\'omez}

\title{Multi-Messenger Constraints on Magnetic Fields in Merging Black Hole-Neutron Star Binaries}

\author[0000-0002-1271-6247]{Daniel J. D'Orazio}
\affiliation{Niels Bohr International Academy, Niels Bohr Institute, Blegdamsvej 17, 2100 Copenhagen, Denmark}
\email{daniel.dorazio@nbi.ku.dk}

\author[0000-0003-3633-5403]{Zolt\'an Haiman}
\affiliation{Department of Astronomy, Columbia University, New York, NY, 10027, USA}

\author[0000-0002-5623-6165]{Janna Levin}
\affiliation{Department of Physics and Astronomy, Barnard College of Columbia University,
New York, New York 10027, USA}

\author[0000-0003-0607-8741]{Johan Samsing}
\affiliation{Niels Bohr International Academy, Niels Bohr Institute, Blegdamsvej 17, 2100 Copenhagen, Denmark}

\author[0000-0003-1817-3586]{Alejandro Vigna-G\'omez}
\affiliation{Niels Bohr International Academy, Niels Bohr Institute, Blegdamsvej 17, 2100 Copenhagen, Denmark}
\affiliation{DARK, Niels Bohr Institute, University of Copenhagen, Jagtvej 128, 2200, Copenhagen, Denmark}

\begin{abstract} 
The LIGO-Virgo-KAGRA Collaboration recently detected gravitational waves (GWs) from the merger of black-hole-neutron-star (BHNS) binary systems GW200105 and GW200115. No coincident electromagnetic (EM) counterparts were detected. While the mass ratio and BH spin in both systems were not sufficient to tidally disrupt the NS outside of the BH event horizon, other, magnetospheric mechanisms for EM emission exist in this regime and depend sensitively on the NS magnetic field strength. Combining GW measurements with EM flux upper limits, we place upper limits on the NS surface magnetic field strength above which magnetospheric emission models would have generated an observable EM counterpart. We consider fireball models powered by the black-hole battery mechanism, where energy is output in gamma-rays over $\lesssim1$~second. Consistency with no detection by Fermi-GBM or INTEGRAL SPI-ACS constrains the NS surface magnetic field to  $\lesssim10^{15}$~G.  Hence, joint GW detection and EM upper limits rule out the theoretical possibility that the NSs in GW200105 and GW200115, and the putative NS in GW190814, retain $\gtrsim10^{15}$~G dipolar magnetic fields until merger. They also rule out formation scenarios where strongly magnetized magnetars quickly merge with BHs. We alternatively rule out operation of the BH-battery powered fireball mechanism in these systems. This is the first multi-messenger constraint on NS magnetic fields in BHNS systems and a novel approach to probe fields at this point in NS evolution. This demonstrates the constraining power that multi-messenger analyses of BHNS mergers have on BHNS formation scenarios, the magnetic-field evolution in NSs, and the physics of BHNS magnetospheric interactions.

\end{abstract}


\section{Introduction}
The LIGO-Virgo-KAGRA (LVK) collaboration recently announced the first definitive detections of gravitational waves (GWs) from neutron-stars merging with black-holes in the black-hole-neutron-star (BHNS) binaries GW200105 and GW200115 \citep{LIGO_NSBHs:2021}. These two events join the suspected BHNS system GW190814 with inferred secondary mass on the cusp between classification as a large NS and a small BH \citep{GW190814:2020}. BHNS systems have long attracted interest for their ability to generate louder GWs than binary NSs (thanks to the BH), while still promising a bright electromagnetic (EM) counterpart (thanks to the NS). Historically, the disruption of the NS has been the focus of possible EM counterparts to BHNS mergers \citep[\eg,][]{NarayanPacz:1992}. However, disruption occurs only for a small region of BHNS parameter space, when the BH is highly spinning and small, $\lesssim 5-10 \Msun$ \citep[see][and \S\ref{sS:Disrupt} below]{Foucart:2018}. Furthermore, given the LVK population of compact object mergers discovered so far, it is likely that most BHNS mergers will not disrupt the NS \citep{FragioneNSBH:2021}.

However, as if often overlooked, the lack of a NS disruption does not imply the lack of a bright EM counterpart, and hope is not lost for BHNS multi-messenger astronomy. Bright EM emission is possible and even expected from non-disrupting systems. The reason lies in the magnetic energy locked up in the NS magnetosphere, which can be tapped without breaking apart the NS just as it is tapped to power the many observational manifestations of known NSs, \eg, the pulsars, anomalous x-ray pulses, soft-gamma repeaters, and others \citep[see, \eg,][]{Kaspi:2010}.

A promising mechanism for brightening non-disrupting BHNS systems arises through a coupling of the orbital and magnetic field energy of the merging binary. Magnetic fields anchored in the NS pass over the BH horizon generating an electromotive force that can supply power to accelerate charges in the magnetosphere analogously to a unipolar inductor (homopolar generator). Such a mechanism has been studied extensively in the literature, originally applied to planetary systems \citep{GLB:1969} and more recently applied to exoplanetary systems \citep{LaineLin:2012}, compact object mergers \citep{Vietri1996, HansenLyut:2001, DLai:2012, Piro:2012} and systems including a BH \citep{McL:2011, Lyut:2011, DL:2013, DL:2016}, in which case we refer to this energy source as the black-hole battery. Encouragingly, numerical investigations of the BHNS magnetosphere have found magnetospheric currents and Poynting flux measurements that agree well with, and possibly exceed, estimates in the above, analytical works \citep{Paschalidis+2013, Carrasco+2021}.

A key feature of BH-battery magnetospheric emission is its strong dependence on the strength and configuration of the large-scale magnetic fields anchored in the NS. The total power available for liberation via the BH-battery scales with the square of the magnetic field strength at the BH horizon. Hence, if the binary parameters are known from GW observations and consistent with a non-disrupting system, then an EM observation consistent with a BH-battery powered event provides information on the NS magnetic field as well as insight into the BHNS magnetospheric physics. Similarly, a GW-confirmed non-disrupting BHNS plus an EM upper limit places an upper limit on the magnetic field anchored in the NS at merger.

Here we use this feature of the predicted BHNS magnetospheric emission, along with GW observations and EM upper limits for LVK events GW200105, GW200115, and GW190814 to make {\em the first multi-messenger constraints of NS magnetic field strengths in BHNS binaries.} To do so we employ a model for the EM signature of a BH-battery powered event developed in \citet{DL:2016} (hereafter \citetalias{DL:2016}) which predicts a short-hard burst of gamma rays within approximately a second after the merger. For each system we find that the strength of a dipolar field at the NS surface is constrained to be $\lesssim 10^{15}$~G, within the astrophysically interesting realm of magnetar field strengths. Future GW detections plus gamma-ray flux upper limits could offer constraints as low as $10^{13-14}$~G depending on source distance, binary parameters, and EM flux upper limits. Given the BHNS event rate, a few events with meaningful flux upper limits are expected per year. This method offers not only a new observational constraint on NS magnetic field evolution but also a handle on BHNS formation channels.

This work is organized as follows. In \S \ref{S:Obs} we provide an overview of the GW observations and EM upper limits. In \S \ref{S:EmissionModels} we present a summary of the BH-battery powered fireball model for non-disrupting magnetospheric emission in BHNS systems. In \S \ref{S:Results} we apply this model to obtain magnetic field strength upper limits for GW200105, GW200115, and GW190814. We discuss astrophysical implications and prospects for the future in \S \ref{S:Discussion} and conclude in \S \ref{S:Conclusion}.

\section{Overview of Observations}
\label{S:Obs}
\subsection{Gravitational Waves}

The GW observations for GW200105 and GW200115 are detailed in \citet{LIGO_NSBHs:2021}. We assume, as concluded in \citet{LIGO_NSBHs:2021}, that each binary consists of a more massive primary BH component and a less massive secondary NS component. For completeness we also consider GW190814 for which association with a BHNS merger is likely but less certain given the large mass of the secondary \citep{GW190814:2020, Antoniadis+2021}. Table \ref{table:GW} lists the relevant binary parameters that we use in this work: BH mass $M_{\BH}$, NS mass $M_{\NS}$, BH dimensionless spin $S_{\BH}$ (ranging from -1,1), and luminosity distance $D_L$ \footnote{To convert between luminosity distance and redshift we assume the same cosmology as \cite{LIGO_NSBHs:2021}, with $h=0.679$, $\Omega_\mathrm{M}=0.3065$.}. For all parameters we quote the median values with $90\%$ credible intervals. For GW2001015 and GW200115 we assume the low-NS-spin prior of \citet{LIGO_NSBHs:2021}.

\begin{table*}
\begin{center}
\begin{tabular}{ l | c | c | c | c } 
 Event    & $M_{\BH}$ [$\Msun$] & $M_{\NS}$ [$\Msun$] & $S_{\BH}$ & $D_L$ [Mpc]  \\
 \hline
 {\bf GW200105} \citep{LIGO_NSBHs:2021} & $8.9^{+1.1}_{-1.3}$ & $1.9^{+0.2}_{-0.2}$ & $0.09^{+0.18}_{-0.08}$ & $280^{+110}_{-110}$ ($z=0.06^{+0.02}_{-0.02}$) \\
 {\bf GW200115} \citep{LIGO_NSBHs:2021} & $5.9^{+1.4}_{-2.1}$ & $1.4^{+0.6}_{-0.2}$ & $0.31^{+0.52}_{-0.29}$ & $310^{+150}_{-110}$ ($z=0.07^{+0.03}_{-0.02}$) 
 \\
  {\bf GW190814} \citep{GW190814:2020} & $23.1^{+1.0}_{-1.1}$ & $2.59^{+0.09}_{-0.08}$ & $\lesssim0.07$ & $241^{+45}_{-41}$ ($z=0.053^{+0.009}_{-0.01}$) 
\end{tabular}
\end{center}
\caption{Double compact object parameters determined from gravitational wave observations.} 
\label{table:GW}
\end{table*}

\subsection{Electromagnetic Upper Limits}
\label{S:ObsEM}

Here and in Table \ref{table:EM} we summarize the significant gamma-ray upper limits and excesses reported for each event as reported in the GCN archives for GW200105 \citep{GCN_105}, GW200115 \citep{GCN_115}, and GW190814 \cite{GCN_190814}.
For all three events, no signal was detected by the Fermi-GBM above the background providing a 3-sigma upper limit weighted by GW localization probability. Upper limits on the gamma-ray flux were placed using the analysis pipeline of \citet{Blackburn+2015}, updated by \citet{O2GBM_searchupdate:2016}. The upper-limit depends on the expected signal duration and spectral template. The standard analysis pipeline considers three spectral templates: `soft', `normal', and `hard', and three different durations: `short' (0.128 s), `medium' (1.024 s), and `long' (8.192 s). We ignore the long-and soft combinations as these best describe long gamma-ray bursts (GRBs) while the BHNS EM counterparts detailed below are closer to hard, short GRBs. The `hard' template assumes a Band-function \citep{Band:1993} with a peak energy at $1$~MeV, and power law slopes $\alpha=0$, $\beta=1.5$, while the `normal' spectrum assumes a peak energy at $0.23$~MeV, and power law slopes $\alpha=1$, $\beta=2.3$ \citep{Blackburn+2015}. For comparison, the energies of emission for the BHNS fireball (detailed below) peak at $2.4 \ \mathrm{MeV} \ \left( B_{\NS}/10^{14} \mathrm{G}\right)^{1/2}$ with approximate durations of $\sim$$0.1 \mathrm{s} \left( B_{\NS}/10^{14} \mathrm{G}\right)$. Hence, we choose to consider a range of upper limits given source templates that range from medium-normal to short-hard, with the short-hard spectral template providing the most conservative upper-limits. The resulting Fermi-GBM (3-sigma) upper-limit ranges are listed in Table \ref{table:EM}.

Table \ref{table:EM} also lists a flux upper limit for GW200105 from INTEGRAL SPI-ACS\footnote{INTEGRAL did report one possibly associated excess with a $\sim 10^{49}$ erg/s/cm$^2$ inferred luminosity, occurring $1.1$~sec before the merger with signal-to-noise ratio of $3.3$ and a false-alarm probability of $0.01$.}. 
This limit is placed using a template in-between the Fermi medium-normal and short-hard templates, with an exponentially cutoff $\alpha=-0.5$ power law and a peak energy at 0.6~Mev, assuming a burst lasting $\leq 1$~s. This is consistent with the range inferred from Fermi.

\begin{table*}
\begin{center}
\begin{tabular}{ l | c | c | c | c } 
          \vspace{-4pt}
          & Obs. Loc. Prob. & Time Window & Energy Range  & Upper Limit (med.-norm. $-$ short-hard) \\
 Detector &[$\%$] & [sec] & [MeV] & $\left[10^{-7} \mathrm{erg/s/cm}^2 \right]$ \\ 
 \hline
& \multicolumn{4}{c}{\textbf{\em GW200105}} \\
\hline
Fermi-GBM & $83.2$ & $t_{\mathrm{mrg}} \pm 30$ & $10^{-2}-1$& $1.8-11$ \\ 
" & "  & " & $10^{-3}-10$& $2.5-25$ \\ 
INTEGRAL SPI-ACS & $50.0$ & $t_{\mathrm{mrg}} \pm 300$ & $7.5\times 10^{-3}-2$ & 2.2 \\
\hline
& \multicolumn{4}{c}{\textbf{\em GW200115}} \\
\hline
Fermi-GBM & $84.4$ & $t_{\mathrm{mrg}} \pm 30$ & $10^{-2}-1$ & $2.8-17$ \\ 
" & " & " & $10^{-3}-10$& $4.0-40$ \\ 
\hline
& \multicolumn{4}{c}{\textbf{\em GW190814}} \\
\hline
Fermi-GBM & $100$ & $t_{\mathrm{mrg}} \pm 30$ & $10^{-2}-1$ & $0.9-5.8$ \\ 
" & " & " & $10^{-3}-10$& $1.3-13$ 
\end{tabular}
\end{center}
\vspace{-10pt}
\caption{Gamma-ray upper limits from GCN archives.}
\label{table:EM}
\end{table*}

\section{Non-Disruption and Emission Model}
\label{S:EmissionModels}

Given the binary properties inferred from GW observations, we will use the EM upper limits and a model for EM emission to place constraints on the NS magnetic field at merger. Hence, we next demonstrate that the NSs in GW200105, GW200115, and GW190814 were very likely not disrupted by their respective BHs and review the model of \citetalias{DL:2016} for BH-battery powered magnetospheric emission in non-disrupting BHNS systems.

\subsection{Disruption or not?}
\label{sS:Disrupt}

\begin{figure}
\begin{center}$
\begin{array}{c}
\includegraphics[scale=0.56]{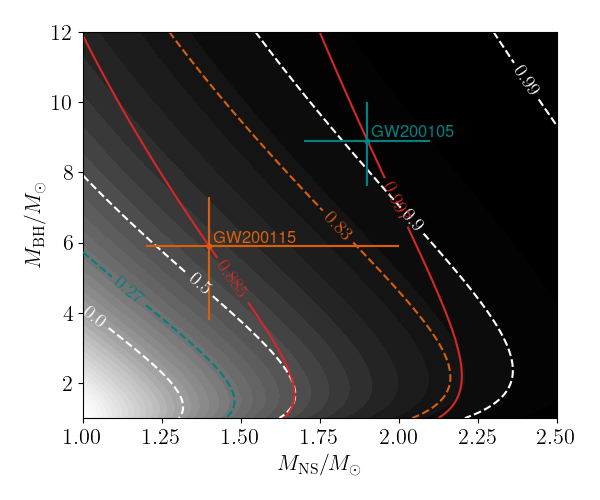}
\end{array}$
\end{center}
\vspace{-20pt}
\caption{
White-to-black contours, with specific values marked by the dashed curves as labeled, represent the maximum aligned component of the BH spin for which a BHNS system can be swallowed whole, given BH mass (y-axis), NS mass (x-axis), and NS radius (fixed at $R_{\NS}=10$~km). The teal (orange) dashed lines represent the 90$\%$ confidence upper limits on the BH spin from GW200105 (GW200115) from Table \ref{table:GW}. For example, systems lying on the white line labeled $0.9$ will be swallowed whole if the orbit-aligned BH spin component is below $0.9$ -- they will at least partially disrupt for aligned BH spin component above $0.9$. Solid red contours represent the required aligned spin component above which a full disruption occurs. Teal (Orange) data points with errorbars represent the values quoted in Table \ref{table:GW}. 
GW190814 would lie to the far upper right, beyond the plotted domain.
}
\label{Fig:Disrpt}
\end{figure}

What portion of allowed binary parameter space results in NS disruption outside of the BH horizon? We focus on GW200105 and GW2100115 as the $23\Msun$, non-spinning BH in GW190814 precludes it from any likely NS disruption. In Figure \ref{Fig:Disrpt}, we use the criteria of \citet{Foucart:2018}, calibrated using General Relativistic simulations of BHNS mergers on circular orbits, to estimate the BH and NS parameters for which NS disruption occurs.
Given BH and NS masses and a choice of $\RNS=10$~km, we solve for a maximum value of the orbit-aligned component of the BH spin, $S_{\BH}$, below which no remnant NS mass is liberated and escapes beyond the BH event horizon. These represent maximum aligned spin components above which at least a partial disruption would occur and are represented in Figure \ref{Fig:Disrpt} by the white-to-black contours, with selected values marked by the dashed curves, as labeled. The two data points and associated errorbars represent the median and $90\%$ confidence uncertainties on the BH and NS masses in GW200105 (teal) and GW200115 (orange; see Table~\ref{table:GW}). In addition to the BH spins below which the NS is swallowed whole, we also solve for the aligned component of the black hole spin above which a full disruption would occur. We draw red contours for these spins at values that pass through the median masses for GW200105 and GW200115.

The dashed teal (orange) contours correspond to the $90\%$ confidence upper bound on the BH spins of GW200105 (GW200115) as inferred from the GW measurements (Table~\ref{table:GW}).  For GW200105, the dashed teal contour demonstrates that the maximum allowed spin magnitude at $90\%$ confidence of $0.27$ is well below the value of $\sim$$0.9$ that is required to disrupt the NS. Hence, no allowed combination of the BH spin and component masses results in even a partial disruption for GW200105. For GW200115, no disruption will occur for the median mass and spin values, but partial disruption can occur for the largest allowed BH spin values (the dashed orange contour at $S_{\mathrm{BH}}=0.83$) and all but the upper limits on the component masses, assuming spin-orbit alignment.  Any misalignment makes the NS even more difficult to disrupt. The red contours show that a full disruption occurs only if $S_{\BH}\gtrsim0.885$ for GW200115, or $S_{\BH}\gtrsim0.997$ for GW200105. Hence, neither system could have been fully disrupted given the $90\%$ confidence system parameters. Hence, a short-GRB-like event due to NS disruption is unlikely.

\subsection{BH-Battery Powered Fireball}

We next review the fireball model of \citetalias{DL:2016}. In this model, a relative velocity between the NS-sourced magnetic fields and the BH horizon generates an electromotive force in analogy to a unipolar inductor \citep[][]{GLB:1969}. The voltage drop corresponding to this BH-battery can be written in terms of binary parameters and the NS magnetic field strength as \citep{McL:2011}
\begin{eqnarray}
V_{\mathcal{H}} = 2 R_{\mathcal{H}} \left[ \frac{\Rbin \left (\Omega_{\orb} - \Omega_{\NS} \right)}{c} + \frac{S_{\BH}}{4\sqrt{2}} \right] B_{\rm NS}
\left(\frac{\RNS}{\Rbin} \right)^3,
\label{Eq:VH}
\end{eqnarray} 
where $r$ is the binary separation, $R_{\mathcal{H}}$ is the spin and mass dependent BH horizon radius, $\Omega_{\orb}= \sqrt{G(\MBH+\MNS)/\Rbin^3}$, $\Omega_{\mathrm{NS}}$ is the NS spin angular frequency, $S_{\BH}$ is the dimensionless BH spin, and we have assumed a dipolar magnetic field with NS surface field strength $B_{\rm NS}$. The power available to the BH-battery is given by Ohm's law, the horizon resistivity $\Res_{\mathcal{H}}=4\pi/c$ and the impedance-matching condition $\Res_{\mathcal{H}}=\Res_{\rm NS}$ \citep[see, \eg, ][]{MPBook},
\begin{equation}
\Lum(\Rbin) = \frac{ 2 V^2_{\mathcal{H}}(\Rbin) }{( \Res_{\mathcal{H}} +
  \Res_{\rm NS} )^2 } \Res_{\rm NS}  \rightarrow  \frac{c}{8\pi}V^2_{\mathcal{H}}(\Rbin),
  \label{Eq:Pow}
\end{equation}
where the prefactor counts emission from each hemisphere over which a full voltage drop occurs.

\citetalias{DL:2016} show that at $\sim$$10 \left( B_{\rm NS}/10^{14} \mathrm{G}\right)$~seconds before merger, the magnetosphere becomes opaque to pair-production, primarily from photon-magnetic-field ($\gamma-B$) interactions. Hence, after this point, the BH-battery energy is injected into a hot pair-plasma. Because of the steep $\sim$$\Rbin^{-4}$ scaling of BH-battery power, the injected power can be evaluated as $\Lum(\Rbin_{\mathrm{mrg}})$, where we take the separation at merger to be $\Rbin_{\mathrm{mrg}}=\RNS + R_{\mathcal{H}}$ (see \S \ref{S:CaveatsFBmodel}). The optically thick pair plasma expands under its own pressure until it reaches a radius $R_{\mathrm{ph}}$ a few $\times \sqrt{B/(10^{14}~\mathrm{G})}$ milliseconds after merger where it becomes optically thin to Thomson electron scattering and emits as a black body with an effective
temperature given approximately by,
\begin{equation}
    T_{\mathrm{eff}} = \left( \frac{ \Lum(\Rbin_\mathrm{mrg}) }{ 4 \pi r^2_0 \sigma_{\mathrm{SB}} } \right)^{1/4}, 
    \label{Eq:KT}
\end{equation}
where $\sigma_{\mathrm{SB}}$ is the Stefan-Boltzmann constant, and the initial injection radius is taken to be $r_0 = \Rbin_\mathrm{mrg}$ to approximate the size of the magnetosphere region surrounding the binary at merger (there is very little dependence on this choice). This effective temperature takes into account the adiabatic cooling of the fluid as it expands to $R_{\mathrm{ph}}$ and the counteracting effect (in the observer's frame) of the Doppler boost from the relativistically expanding fireball (see below and \citetalias{DL:2016}).

The resulting spectrum is found from integrating the multi-component black-body emission over a sphere of radius $R_{\mathrm{ph}}$ that is expanding with Lorentz factor $\gamma = R_{\mathrm{ph}}/r_0$ and with rest frame photosphere temperature $T_0/\gamma$, for initial injection temperature $T_0$.  This calculation is carried out in full in \citetalias{DL:2016}, however, for simplicity here we remove the integral over the sphere and approximate the observed flux with the most significantly contributing portion of the emitting photosphere, the portion within half opening angle $\gamma$ and a constant observed temperature at redshift $z$ of $\gamma T_{\mathrm{ph}}(1+z)^{-1} = \gamma \left(T_0/\gamma\right)(1+z)^{-1} = T_0(1+z)^{-1}$ \citep[see, \eg, ][and Eq.~\ref{Eq:KT}]{Pacz:1986GRB},
\begin{widetext}
\begin{equation}
    F_{\mathrm{FB},\obs} \approx  2 \pi \mathcal{K}\left(1 - \cos\gamma^{-1} \right) \left( \frac{ R_{\mathrm{ph}}}{D_L} \right)^2 \int^{\nu_{\rm{max}}}_{\nu_{\rm{min}}}{  \frac{2 h \nu^3}{ c^2} \frac{ d \nu }{\rm{exp\left[\frac{h \nu}{ k T_0}\right] - 1} }  } ,
    \label{Eq:Fobs}
\end{equation}
\end{widetext}
which is the flux from a blackbody with temperature $T_0$ at redshift $z$, corresponding to luminosity distance $D_L$\footnote{Written in terms of angular diameter distance and $T_0(1+z)^{-1}$ in \citetalias{DL:2016}.}. Here $\nu$ is the frequency in the observer's frame, $2 \pi \left(1 - \cos\gamma^{-1} \right)$ is a geometrical factor accounting for the significantly boosted region of the sphere, and $\mathcal{K}$ is a numerical factor of order unity that corrects the single-temperature approximation that allowed removal of the integral over the hemisphere \footnote{Under this single-temperature approximation the spectrum is blackbody but in the full calculation, deviations from blackbody arise because the observed temperature varies across the surface of the photosphere due to the varying line of sight expansion velocity \citepalias{DL:2016}.}. For magnetic field strengths between $10^{12}$ and $10^{16}$, and binary parameters considered here, $\mathcal{K}\approx1.6$ to within a few $\%$.

\begin{figure*}
\vspace{-10pt}
\begin{center}$
\begin{array}{cc}
\includegraphics[scale=0.5]{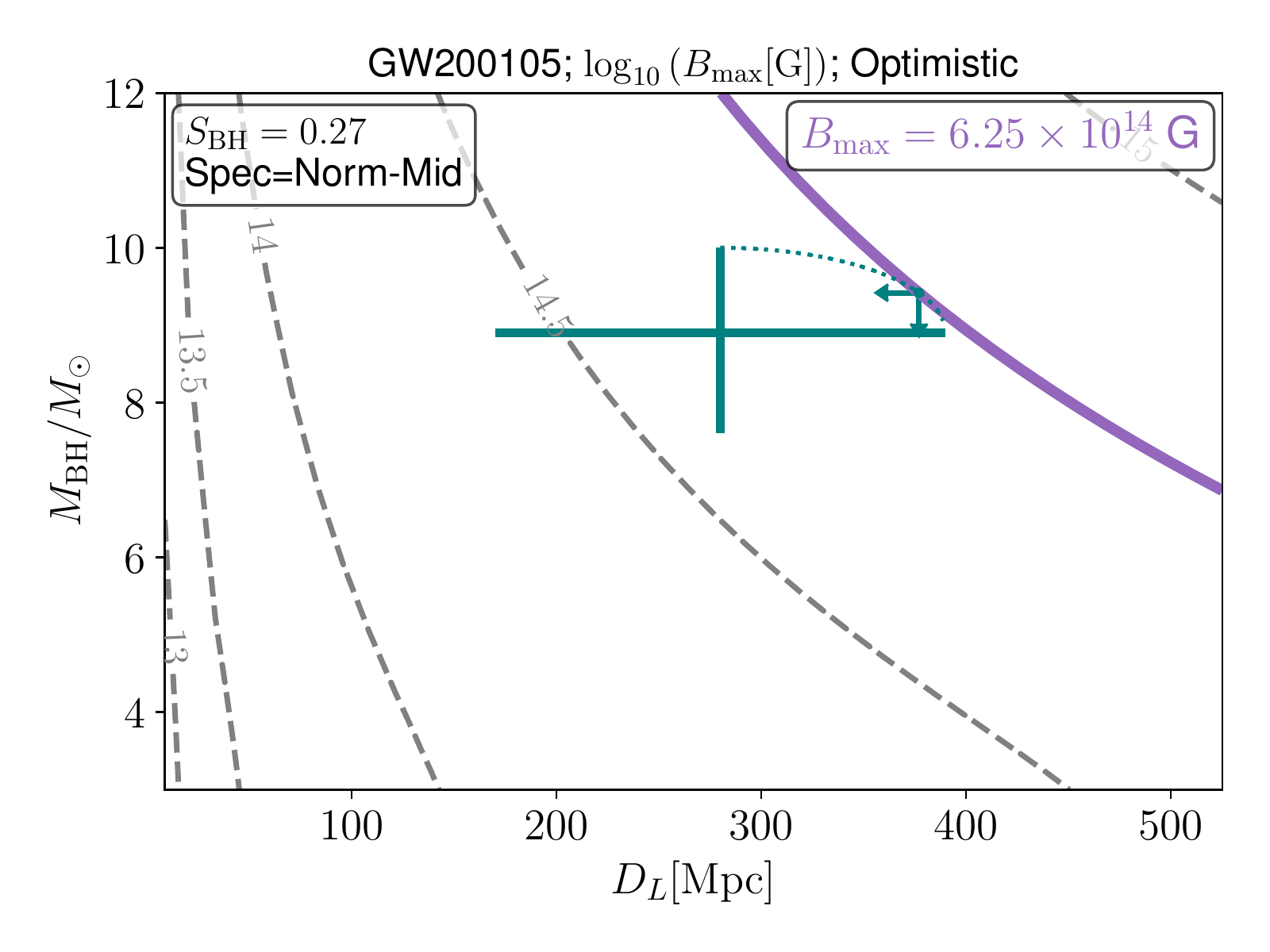} &
\includegraphics[scale=0.5]{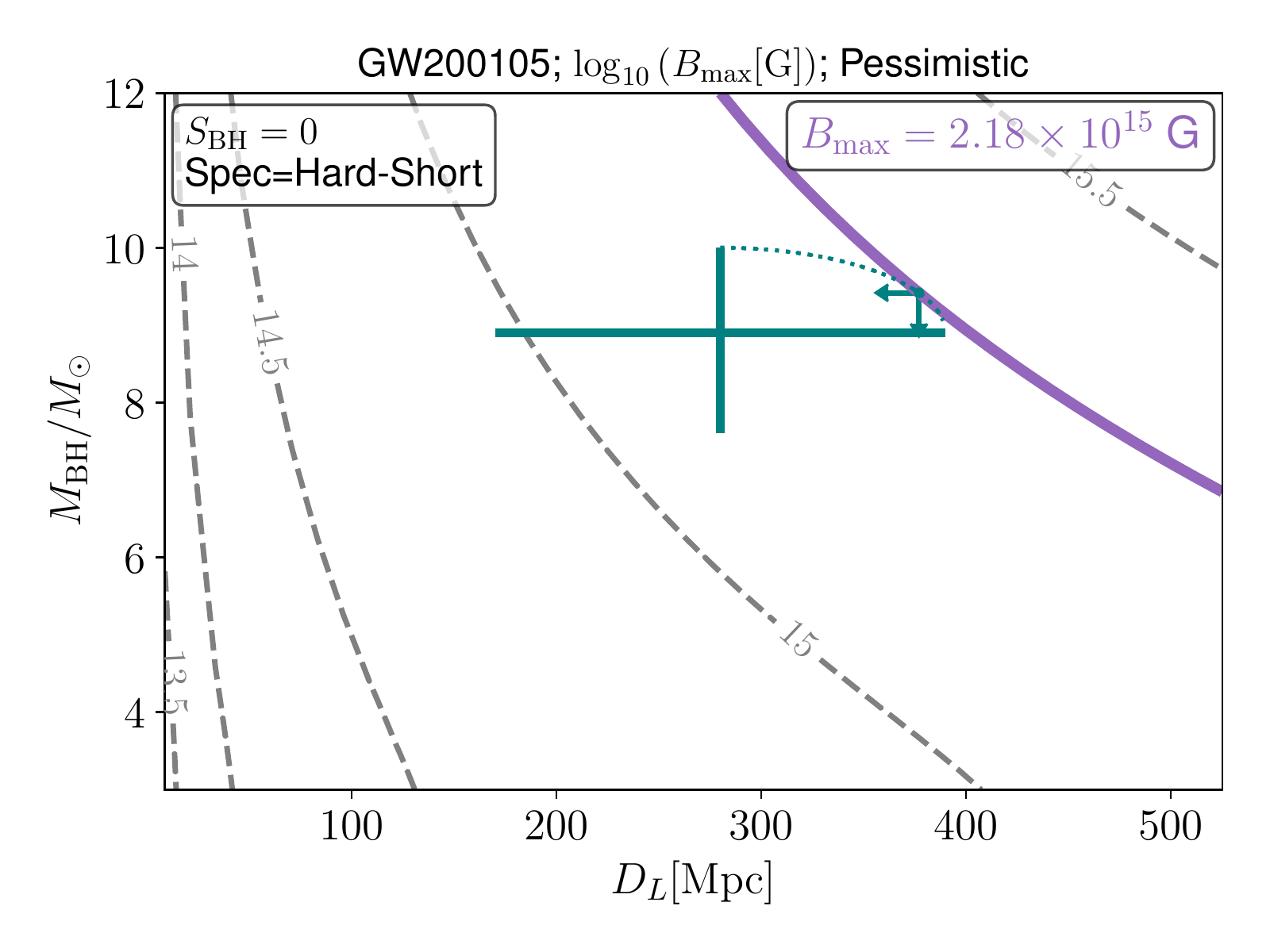} \vspace{-23pt} \\
\includegraphics[scale=0.5]{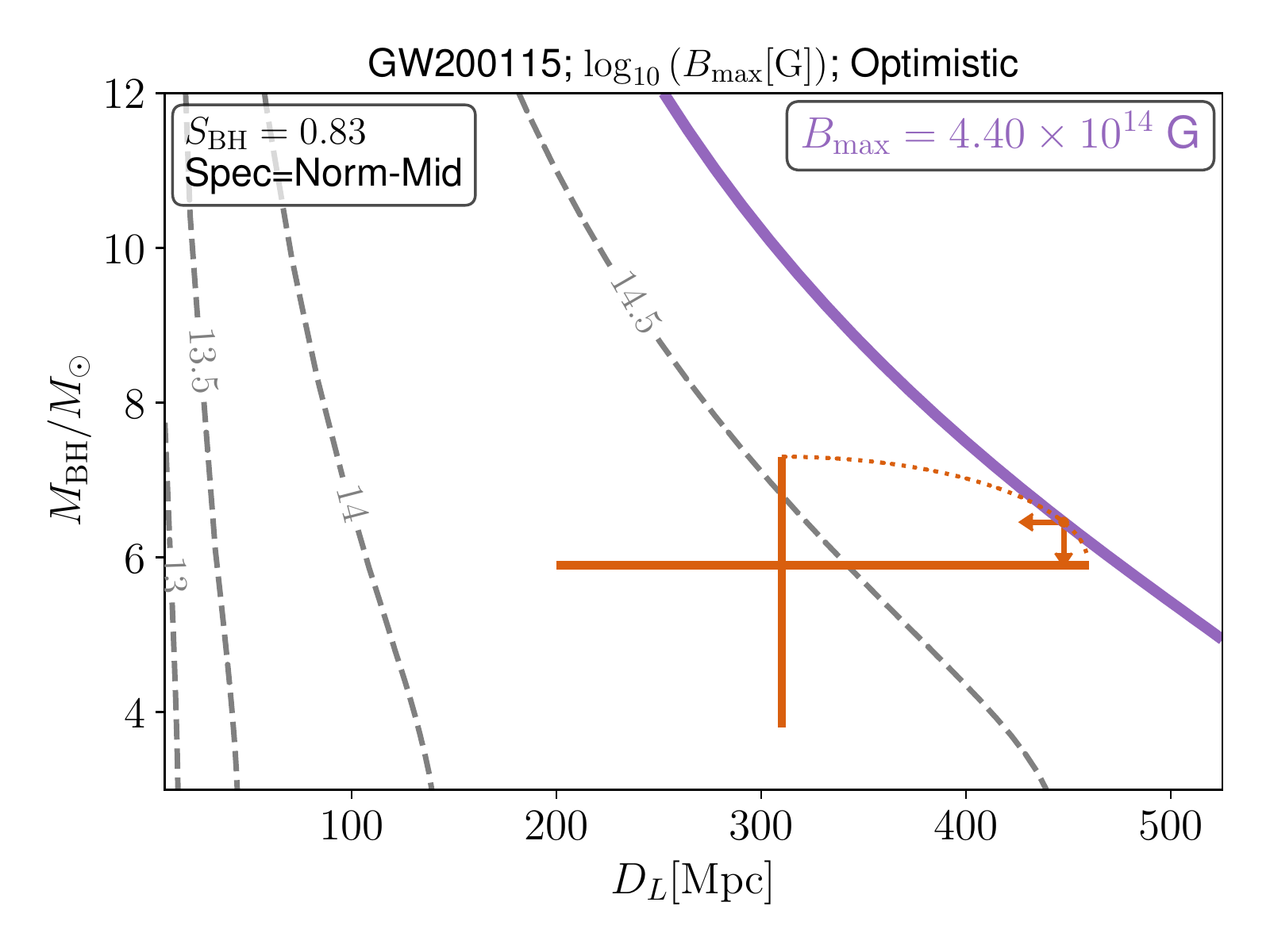} & 
\includegraphics[scale=0.5]{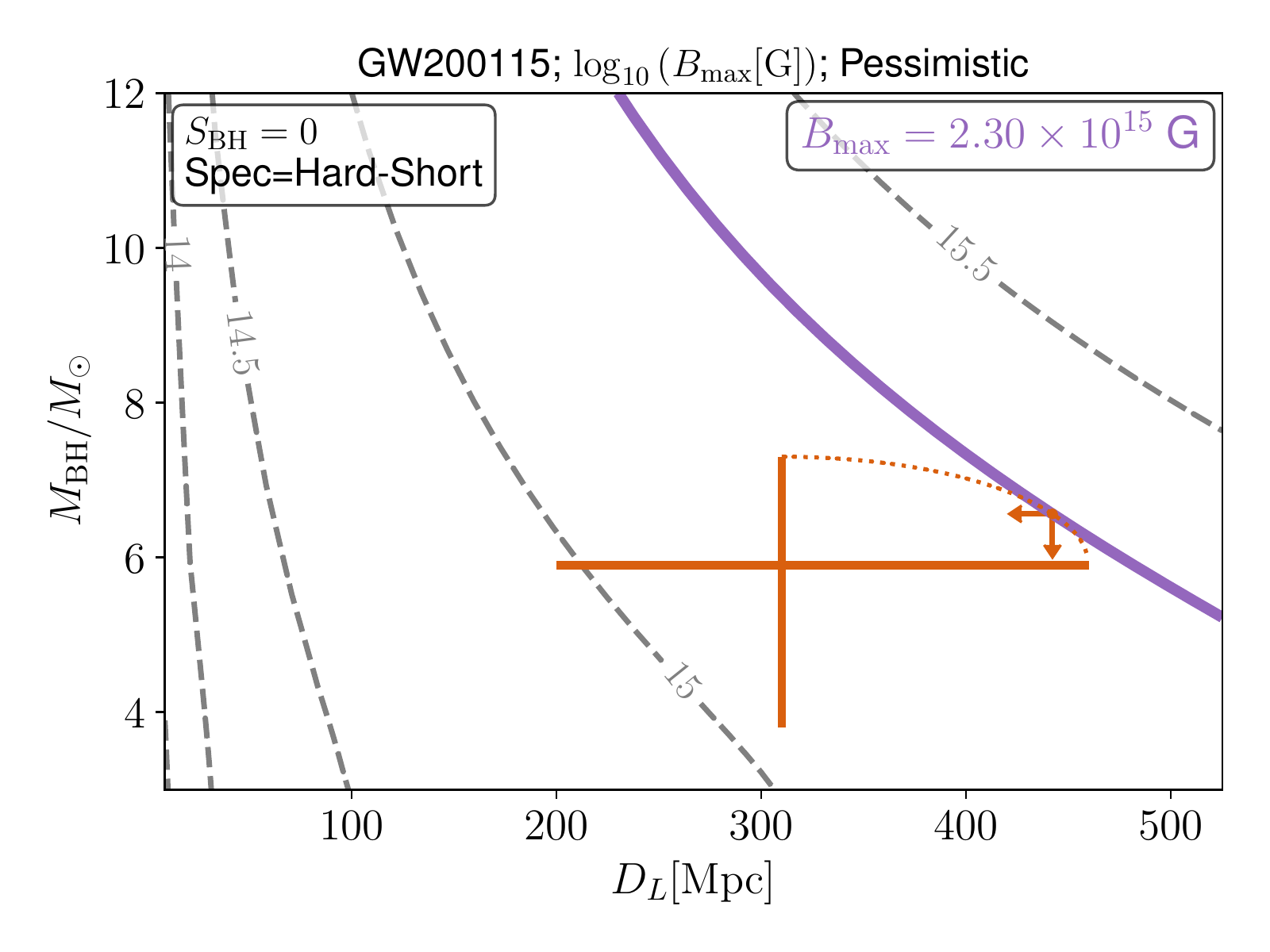} \vspace{-23pt} \\
\includegraphics[scale=0.5]{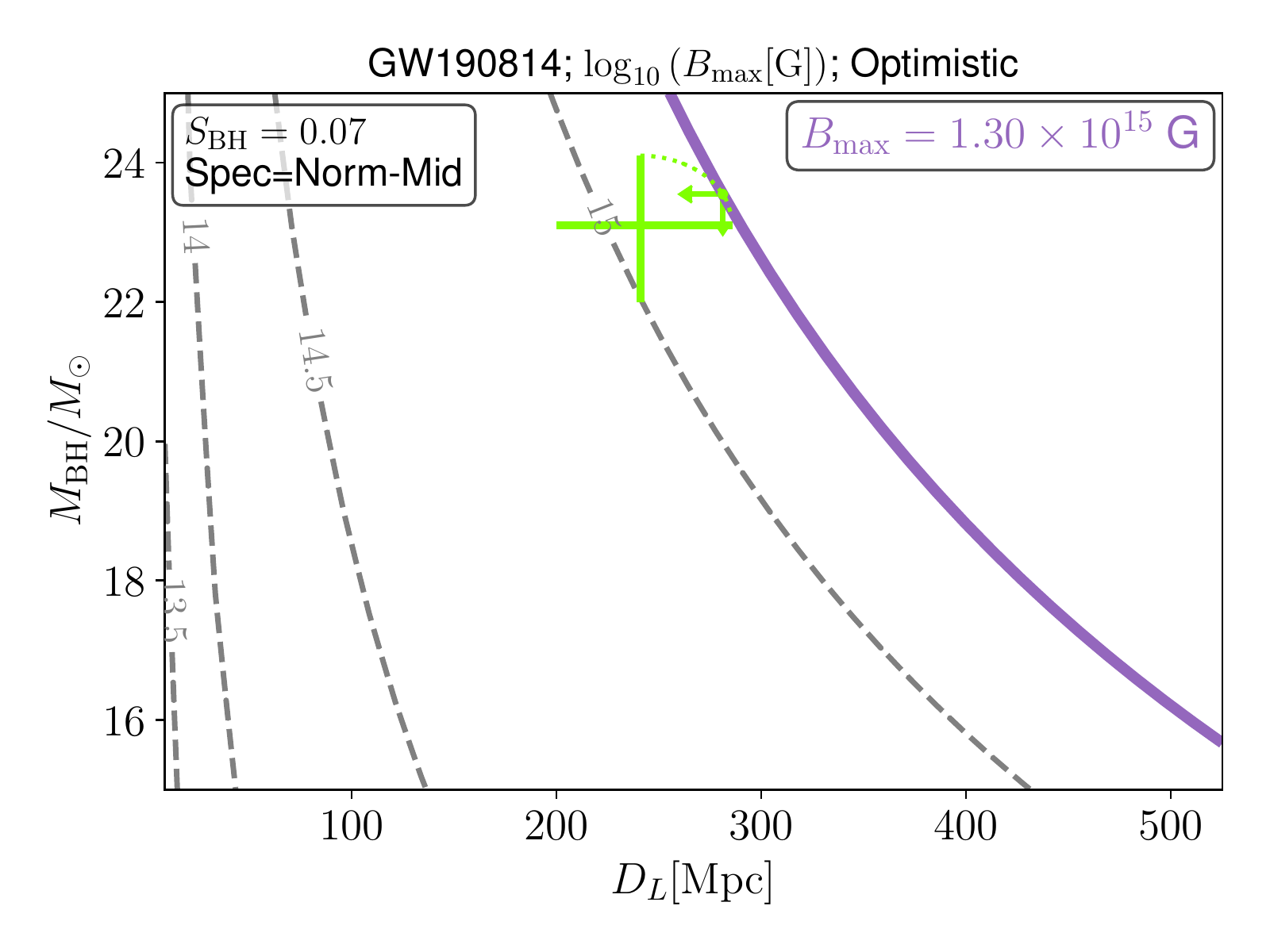} & 
\includegraphics[scale=0.5]{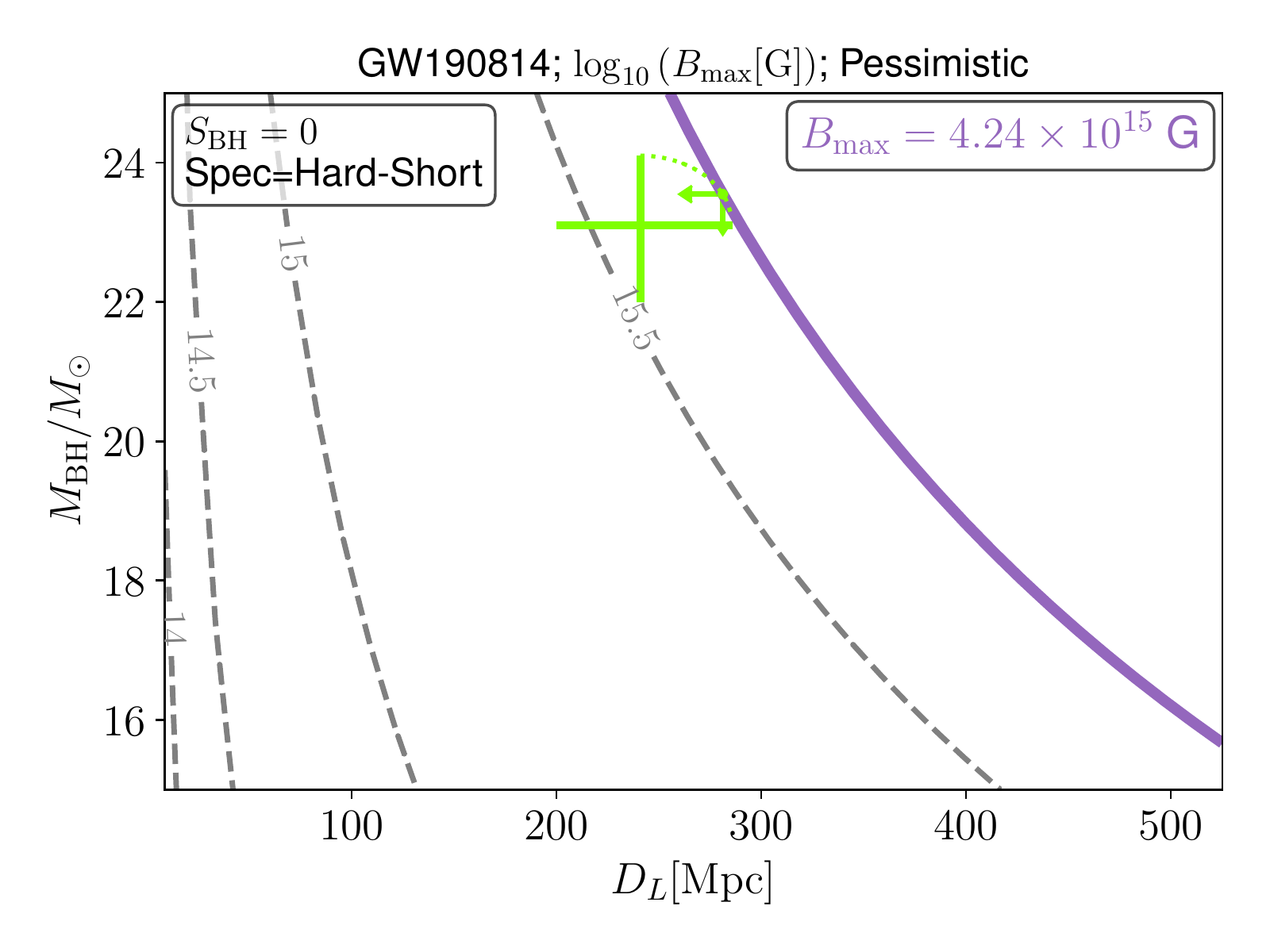}
\end{array}$
\end{center}
\vspace{-23pt}
\caption{
NS surface magnetic-field-strength constraints for GW200105 (top), GW200115 (middle), and  GW190814 (bottom). 
Grey-dashed contours represent the magnetic field strength required to generate a signal at the Fermi-GBM upper limit in the $10^{-3}-10$ MeV range.  
The dotted lines connecting $\MBH-D_L$ error bars represent the $90\%$ confidence contours for a bivariate Gaussian distribution approximating the joint $\MBH-D_L$ posteriors. The fiducial NS parameter values are the median $M_{\NS}$ from Table \ref{table:GW}, $R_{\NS}=10$~km, and $\Omega_{\NS}=0$. The thick-purple contour is the corresponding $90\%$ confidence upper limit on the magnetic field strength. In each panel, the arrows point in the direction of allowed magnetic field strength. The left column represents an optimistic estimate and uses the medium-normal template spectrum and the maximum value of BH spin allowed from the GW observations. The right column represents a more conservative estimate and uses the short-hard template spectrum and zero BH spin. The two scenarios bracket the range of magnetic field strength upper limits, $B_{\max}$ in Table~\ref{table:Bmax}.
}
\label{Fig:Bcnstrnt}
\end{figure*}

\section{Results}
\label{S:Results}

The total flux predicted within a specified spectral range is determined by the parameters $(B_{\NS}, M_{\NS}, R_{\NS}, \Omega_{\NS}, M_{\BH}, S_{\BH}, D_L)$. For the NS radius we choose $R_{\NS}=10$~km. For the unconstrained NS spin we conservatively choose zero spin as this is likely consistent with astrophysical channels discussed in \S \ref{sS:AstroImp}, which may bring a NS with strong fields to merger. The results are not strongly dependent on the NS spin unless it matches the orbital frequency of the binary (see Eq.~\ref{Eq:VH}). If this occurs in the retrograde sense, then the available power increases by a factor of four. In the prograde, i.e. tidally locked case, the BH power drops significantly, however, tidal locking is not expected for these systems \citep{BildstenCutler:1992, DLai:2012}.

Provided these NS radius and spin parameters, a flux upper limit $F_{\mathrm{UL}}$ over frequency range $\left[\nu_{\min},\nu_{\max}\right]$, and with the GW-measured ranges on $(M_{\NS}, M_{\BH}, S_{\BH}, D_L)$, we solve $F_{\mathrm{UL}} \leq F_{\mathrm{FB},\obs}(B_{\max})$ for the maximum allowed NS surface magnetic field strengths. The most significant constraints come from the $10^{-3}-10$~MeV flux upper limits as these extend to the high energies where the fireball luminosity peaks. Hence, we use the corresponding Fermi-GBM flux upper limits in Table \ref{table:EM} and choose $h\nu_{\min}=10^{-3}$~MeV and $h\nu_{\max}=10$~MeV in Eq. (\ref{Eq:Fobs}).

Figure~\ref{Fig:Bcnstrnt} illustrates the results of this calculation in $M_{\BH}$ vs $D_L$ space as these parameters most greatly affect the result given their uncertainties and dependence in the flux calculation. Each row in Figure~\ref{Fig:Bcnstrnt} is for a different GW source, and the columns bracket the range of optimistic and pessimistic choices for flux upper limit and BH spin.
The higher the prograde spin, the closer together the pair gets at merger, and the higher the power input (Eq.~\ref{Eq:VH}). Hence, the highest spins allowed by the GW observation are used in the optimistic cases, while we choose zero spin in the pessimistic case (consistent with the lowest observationally allowed BH spins). In each panel of Figure~\ref{Fig:Bcnstrnt} we plot (dashed-grey) contours of $\log_{10} B_{\NS}$ required to generate an EM signal with the indicated choice of flux upper limit and BH spin. For each panel we use the median value of $M_{\NS}$ since the allowed range quoted in Table \ref{table:GW} only affects our result for $B_{\max}$ at the percent level.

To place the upper limit $B_{\max}$, we approximate the 2D posterior on $M_{\BH}$ and $D_L$ as a bivariate Gaussian. The relevant $90\%$ confidence contours are plotted as the dotted quarter-ellipses in Figure \ref{Fig:Bcnstrnt}. We then find the values of $(M_{\BH}, D_L)$ which maximise $B_{\max}$ along this ellipse giving the $90\%$ confidence upper limits for $B_{\max}$. This value is denoted in each panel of Figure \ref{Fig:Bcnstrnt} by the thick purple contour. From the pessimistic and optimistic values for each GW source we generate a range of inferred maximum magnetic field values and list these in Table \ref{table:Bmax}. The majority of the difference between the optimistic (opt.) and pessimistic (pes.) results in Table \ref{table:Bmax} is due to the change in the flux upper limit (factor of $\sim$$\sqrt{10}$) as opposed to the change in spin which contributes at most a factor of $\sim$$1.6$ in the case of GW200115, where the spin ranges from $0.0-0.83$.

\begin{table}
\begin{center}
\begin{tabular}{ l  | c } 
           &  $B_{\max}$ [G]   (opt. $-$ pes.)   \\
 \hline
{\bf GW200105} & $6.3 \times 10^{14} -  2.2 \times 10^{15}$ \\
{\bf GW200115} & $4.4 \times 10^{14} - 2.3 \times 10^{15} $ \\
{\bf GW190814} & $ 1.3 \times 10^{15} - 4.2 \times 10^{15} $
\end{tabular}
\end{center}
\caption{Range of magnetic field upper limits $B_{\max}$ using Fermi-GBM upper limits over $10^{-3}-10$ MeV. Optimistic (opt.) and pessimistic (pes.) values bracket the range as described in Fig.~\ref{Fig:Bcnstrnt}.} 
\label{table:Bmax}
\end{table}

\begin{figure}
\begin{center}$
\begin{array}{cc}
\includegraphics[scale=0.53]{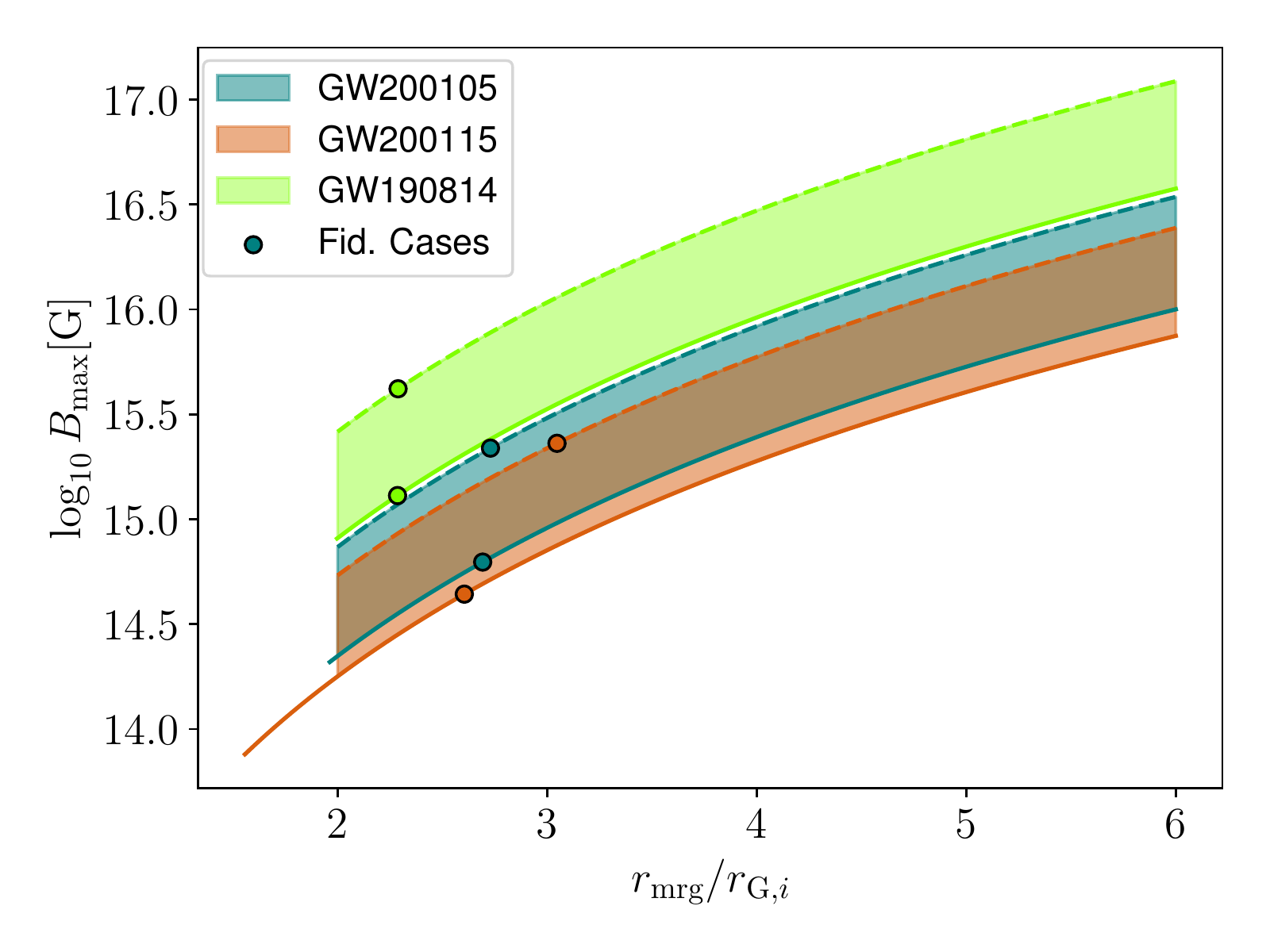}
\end{array}$
\end{center}
\vspace{-20pt}
\caption{
Variation in NS magnetic field upper limit as a function of model uncertainties. The x-axis varies the value of the separation where we assume that the BH-battery power shuts off and the fireball begins expanding (see Eq. \ref{Eq:KT}). The dashed lines correspond to the pessimistic case where $S_{\BH}=0$, and the short-hard flux upper limit is used. The solid lines correspond to the optimistic case, which employs $90\%$ confidence upper limits on the BH spin and the medium-normal flux upper limits. The dots represent the fiducial cases as plotted in Figure~\ref{Fig:Bcnstrnt} and quoted in Table \ref{table:Bmax}. 
}
\label{Fig:rm_var}
\end{figure}

\section{Discussion}
\label{S:Discussion}

\subsection{Astrophysical Implications}
\label{sS:AstroImp}

We have shown that GW observations of BHNS binaries plus EM upper limits can constrain the strength of the magnetic field anchored to the NS as it takes its final plunge across the BH horizon. For the first two definitive BHNS binary detections, an upper limit on the maximum NS field strength can be conservatively placed at $\sim$$ 10^{15}$~G, and slightly higher for GW190814. These values are at the high end of observed magnetar magnetic field strengths and within theoretically allowed values. We now assess the astrophysical implications of these and plausible future upper limits.

\subsubsection{Long-Lived or Rejuvenated Fields}
Theoretical arguments depending on hydro-magnetic instabilities that grow the field during the proto-NS stage predict maximum initial fields strengths up to $\sim5\times10^{16}$~G \citep{Miralles+2002}. Theoretical work on the coupled crust-core magnetic-field evolution in NSs is still under development, but current uncertainty in our understanding leaves open the possibility that large-scale magnetic fields anchored in the superconducting core could persist indefinitely \citep{LyutikovMckinney:2011, Elfritz+2016, BransgroveLevinBelo_2018, Gusakov+2020}. Furthermore, while elegantly motivated by the picture of NS unification \citep{Kaspi:2010}, there is yet no direct evidence for large-scale dipole-field decay in NSs \citep[see][for an overview of existing indirect evidence]{Igoshev+2021}. Hence, the theoretical possibility of long-lived, strong magnetic fields anchored in NS cores combined with a lack of direct evidence for fast dipolar field decay suggests the possibility that NSs could be born with and retain $\lesssim 10^{16}$~G poloidal magnetic field components that they take with them to merger. Indeed the only way to release the energy in these core-anchored fields and so provide evidence for their existence would be for a resulting NS collision with another compact object, either through disruption, or through tapping these fields with the black-hole battery.

{\em Hence, the multi-messenger constraints considered here rule out the existence of long-lived $\gtrsim 10^{15}$~G dipolar magnetic-fields retained on the merging NSs in GW200105 and GW200115.} An estimate of the fraction of NSs expected to be born with fields above $10^{15}$~G, however, is beyond the scope of this work. Similarly, we rule out any mechanism that pumps the large scale poloidal magnetic-field component to high values during inspiral\footnote{See for example the mechanism in \cite{ZrakeMacFad:2013}, which, however, generates a turbulent not and an ordered field, and likely requires full NS disruption.}.

\subsubsection{Magnetar Plus Black Hole Merger Scenarios}

Taking a more conservative direction, we can employ the existence of the observed population of high-magnetic-field NSs, also known as magnetars.
Magnetars have field strengths up to a few times $10^{15}$~G, inferred from spin down arguments \citep[see][and references therein]{Igoshev+2021} and also more direct cyclotron measurements \citep{Tiengo+2013}. From a combination of arguments involving kinematic association with a supernova remnant or young star cluster, NS galactic position, or NS spin, one can infer typical magnetar lifetimes of $\lesssim 10^5$~yrs \citep[see][and references therein]{Igoshev+2021}. Hence, under the assumption that the magnetic field components responsible for magnetar behavior disappear at the end of the magnetar lifetime, and that these are the large scale fields that power the black-hole battery,
{\em then the upper limits derived here rule out GW200105 or GW200115 as products of a (strong-field)-magnetar-BH merger.} A necessary condition for such a merger is that the time between magnetar formation and merger is shorter than the magnetar lifetime of $10^4-10^5$~yrs. It is then natural to ask what formation channels, if any, can act on this timescale and can so be ruled out for these and similar future events.

\paragraph{Dynamical Pairing}

If a magnetar is formed in a dense cluster, such as a nuclear star cluster, it can be brought to merge with a BH through various dynamical pathways. One is through direct GW capture, where the initially unbound NS becomes bound to a BH through the emission of GWs at its first pericenter passage. Considering now a system with constant velocity dispersion, $v$, and uniform density of BHs, $n$, the characteristic time for a NS to be captured by a BH is $t_c \sim 1/(n \sigma_{c} v)$, where $\sigma_{c}$ is the capture cross section \cite[\eg,][]{1993ApJ...418..147L}. For a population of BHs with mass $M = 10 M_{\odot}$, and a NS with mass $m = 1.4 M_{\odot}$ one then finds $t_c \approx 10^{12}\ \mathrm{yrs} \times \ (v/(10\ \mathrm{kms}^{-1}))^{11/7} (n/10^5 \mathrm{pc}^{-3})^{-1}$. Therefore, pairing up magnetars through GW capture chance encounters in dense clusters is highly unlikely. The same conclusion, but with different scalings, is found in the case the merger is mediated through a 3-body interaction between a binary BH and the NS, which suggests that BH-magnetar mergers are not easily assembled through standard chaotic dynamics, unless more exotic pathways are considered (see, \eg, the active galactic nuclei scenario \citealt{McKernanNSBH_AGN:2020, Yang_Gayathri+2020, TagawaKocsis+2021}).

That the merger rate involving NSs in stellar clusters generally is low \citep{2021arXiv210714239M} has also recently been shown using state-of-the-art Monte Carlo cluster codes \citep{Ye_GCNSBH+2020}.

\paragraph{Binary Stellar Evolution}
The merger rates of BHNS binaries are predicted to be dominated by massive binaries that evolved in isolation in contrast to those that evolve in dynamical environments \citep[see][for a review]{2021arXiv210714239M}. Therefore, we use a synthetic population created with the rapid population synthesis element of the COMPAS \citep{2021arXiv210910352T,2017NatCo...814906S,2018MNRAS.481.4009V} suite v02.19.01 \cite[available via][]{vigna_gomez_alejandro_2021_4682798} to estimate the merger time distribution of BHNSs from isolated binary evolution. The population consists of $10^7$ massive binaries ($M>5 M_{\odot}$) at representative metallicities $Z=\{0.02,0.0142,0.01,0.001,0.0001\}$ following the setup from \cite{2020PASA...37...38V}. We only consider BHNS binaries in which the NS is the second compact object to form.

The number of BHNS binaries per total mass evolved, the yield, is approximately $10^{-5} M_{\odot}^{-1}$ for all metallicities. The fraction of BHNSs with merger times $<0.1$ Myr is $\lesssim 2/1000$. In this population, short-lived BHNS binaries are the product of exotic formation channels and/or fortuitous natal kicks. This fraction rises to $\lesssim 1/100$ for merger times below a few tens of Myr, and $\lesssim 1/10$ for for merger times below a few hundreds of Myr. 

Hence, if we assume that magnetars can be formed in binary systems \citep{MagnetarsinXRBs:2021} and that the NS can only hold onto large magnetic fields for the magnetar lifetime, then $\sim$$10^3$ BHNS mergers formed through isolated binary evolution are required within a detectable EM horizon before the signal described here is expected. On the other hand, we can use this calculation to estimate that after observing $\sim$$10$ BHNS mergers via GWs within a detectable EM horizon and with no EM counterpart, we begin to constrain large magnetic field retention timescales to less than $\sim$$10^8$~yrs. Below we estimate how many such events within a detectable EM horizon are expected with present-day instruments.

Note that both formation scenarios above assume that the magnetar lifetime spans a continuous period starting from NS formation, but there is still much to learn about magnetar life-cycles. There is evidence for evolution into and out of the magnetar stage suggested by some radio pulsars for which a growing poloidal magnetic field is inferred via measurements of the breaking index. For example, at its current rate of inferred field growth, PSR 1734-3333 would be reclassified as a magnetar in a few tens of thousands of years \citep{EspinozaLyne+2011}, suggesting the possibility of a bidirectional evolution between the two classes. Additionally, magnetic fields can be buried by gas accretion and re-emerge later, suggesting the possibility of ``hidden magnetars" \cite{HiddenMagnetars:1999}. These possibilities could increase the chances for magnetar-BH mergers by delaying magnetar field strengths to a later time after magnetar formation.

\subsection{Prospects for future observations}
In the future, when we have $N$ total constraints similar to those considered here, we will be able to establish that $\lesssim 1/N$ BHNS mergers do not, for example, contain NSs with long-lived dipole fields above a given strength. Above we estimated what this fraction might be for some viable astrophysical scenarios; here we estimate what $N$ may grow to given current merger rates and EM horizons.

\paragraph{How many do we expect to see in coming years?} 
    Assuming that these two events are representative of the population, the BHNS merger rate is estimated at $45^{+75}_{-33}$~Gpc$^{-3}$yr$^{-1}$, or higher $130^{+112}_{-69}$~Gpc$^{-3}$yr$^{-1}$ when assuming a broader mass distribution \citep{LIGO_NSBHs:2021}. Considering that the majority of the BHNS parameter space represented in Figure \ref{Fig:Disrpt} does not result in NS disruption, we assume that most of the coming BHNS events will not result in a NS disruption \citep{FragioneNSBH:2021} and provide additional upper limits on the NS magnetic fields strength (if not an actual EM counterpart). Assuming that useful limits can be made for systems out to 300 Mpc, we expect $1.2^{+2.0}_{-0.9}$ or $3.5^{+3.0}_{-1.9}$ events per year of observing time. For even more useful, closer events within 100 Mpc, we expect $0.05^{+0.08}_{-0.03}$ or $0.13^{+0.11}_{-0.07}$ events after one year of observations. Hence a few events similar to those studied here are expected in a coming year of observations and ten years may be needed to detect a sub-100~Mpc system.
    
\paragraph{What field strengths can we expect to probe?}
    From Figure \ref{Fig:Bcnstrnt} we see that moving GW2001105 or GW2001115 closer, to within 100~Mpc can decrease the upper limit to $\sim$$10^{14}$~G. Because the magnetic field strength upper limit scales as the square root of the flux upper limit, a future BHNS merger with similar properties as GW2001105 could result in a $10^{13}$~G constraint for a $100\times$ better flux upper limit than used here. More robust upper limits may be possible in the future not only with more sensitive EM detectors but with GW pre-warnings from low-frequency detectors such as LISA \citep{Sesana:LISALIGO:2016, LISA:2017}, TianQin \citep{TianQin:2016, TianQinMB:2020}, or DECIGO \citep{DECIGO_MB:2018}.

\subsection{Caveats}

\subsubsection{Observational Constraints}
Recently \citet{MandelSmith:2021} carried out a re-analysis of the GW200115 event, specifically on the recovery of mass and spins using a more astrophysically informed prior on the BH spin. \citet{MandelSmith:2021} recover a BH spin that is consistent with zero and a tighter constraint on the BH mass of $7.0^{+0.4}_{-0.4} \Msun$. This lower high-end of the BH mass estimate results in a decreased $B_{\max}$ in the pessimistic zero spin case (see Figure \ref{Fig:Bcnstrnt}). However, in the optimistic maximum spin case, reducing the BH spin from $0.83$ to $0.0$ has the overall effect of increasing $B_{\max}$ for GW200115. This results in a $B_{\max}$ range for GW200115 of $8.1 \times 10^{14}~\mathrm{G} \lesssim B_{\max} \lesssim 2.6 \times 10^{15}~\mathrm{G}$ when using the binary component values inferred in \citet{MandelSmith:2021}.

Throughout, we have used the low-NS-spin prior for the GW parameter values. If we instead use the high-spin priors of \citet{LIGO_NSBHs:2021}, the largest change in the recovered system parameters is at the $\sim$0.1$\Msun$ level for the BH mass, resulting in an insignificant change to our magnetic field upper limits. A larger, though still minimal effect comes from including a non-zero $\Omega_{\mathrm{NS}}$, as discussed in \S \ref{S:Results}.

As discussed in \S \ref{S:ObsEM}, the gamma-ray flux upper limits are derived from temporal-spectral models of the expected emission, based on gamma-ray bursts. We used upper limits that derived from choices of the spectrum and signal duration which most closely match the fireball emission models. These are, however, not tailored to the fireball models. Given uncertainties in the fireball model that we discuss below, however, we expect that the chosen upper limits are sufficient for the level of approximation here.

\subsubsection{Fireball Model}
\label{S:CaveatsFBmodel}
Throughout we have assumed that the total energy injected into the fireball is the full BH-battery power at a separation $r_{\mathrm{mrg}} = R_{\NS} + R_{\mathcal{H}}$. This is motivated by the final binary separation where magnetic fields from the NS can be dragged across the BH horizon and generate the voltage of Eq. (\ref{Eq:VH}). Figure~\ref{Fig:rm_var} shows how our results for the magnetic field upper limits varies for a range of choices for $r_{\mathrm{mrg}}$. For the largest $r_{\mathrm{mrg}}$, we conservatively choose the ISCO of a test mass orbiting a non-spinning BH. For the smallest $r_{\mathrm{mrg}}$, we choose the natural scale of the problem, the horizon radius. At such small separations one might expect that the BH-battery operation as we describe it here breaks down or that power is swallowed by the BH. Meanwhile, the larger, non-spinning ISCO separations may be too conservative as the BH-battery may not yet be out of power at this earlier point in the merger. We make our fiducial choice (denoted by the dots in Fig.~\ref{Fig:rm_var}) because even if the BHNS is no longer on adiabatically decaying circular orbits below the ISCO (due also to tidal and higher order relativistic effects), it still has relativistic motion that continues to generate an electromotive force and power the BH-battery \citep[\eg,][]{DL:2013}. For highly spinning BHs, the ISCO approaches $R_{\mathcal{H}}$ and is smaller than our fiducial $r_{\mathrm{mrg}}$. Further understanding of the BH-battery powered emission may eliminate the need to include this parameter or help to choose its value more accurately.

We have not considered mis-alignment of the BH spin, NS magnetic field, and orbital angular momentum \citep[however, see][]{Carrasco+2021} nor have we considered non-dipolar magnetic field configurations. We have not considered beaming of the signal, which would increase the volume to which events could be observed, while decreasing the fraction of BHNS systems for which magnetospheric emission would be observed. Thus far, numerical force-free General-Relativistic simulations of BHNS magnetospheres predict only weak emission anisotropy -- \cite{Paschalidis+2013} find that the Poynting flux carries away energy in a broad beam that could result in a lighthouse-like effect.

\vspace{20pt}

\subsubsection{Non-Fireball Emission}
We have focused on magnetospheric emission via the \citetalias{DL:2016} fireball model because it relies on relatively simple, robust energetics arguments and predicts a frequency dependent flux with which we can compare with observed upper limits. However, just as the escaping radiation can take on many forms in the analogous pulsar magnetosphere so could it for the BHNS binary magnetosphere. For example, \citet{Mingarelli+2017} argue that the BH-battery could power radio emission reminiscent of fast radio bursts. \citet{Carrasco+2021} carry out global force-free simulations of a NS with a pure dipole magnetic field orbiting in a Kerr spacetime finding $\sim$$10\times$ BH-battery luminosities and finding that flux tubes transport electric currents similarly to what is predicted in the analytical BH-battery models (see also \citealt{Paschalidis+2013}), finding additionally that energy can be transported outwards by re-connection and large scale Alfven waves. They consider EM signatures at radio frequencies, but other manifestations of this energy transport are possible and require further study. A number of recent works have also begun to develop possible signatures due to charging and discharging of the BH via the \citet{Wald:1974} mechanism \citep{ChenDai:2021,LDG:2018,Dai_NSBHQ:2019,BZang_ChargedEMC:2019,PanYang_BHdischarge:2019,ZhongDD+2019,Yang+2020}. Interestingly many of these mechanisms still have total power that scales with the square of the magnetic field strength, but expected energies of emission will likely vary.
Crustal cracking in the NS may allow for an EM counterpart which is also strongly dependent on magnetic field strength and could be considered alongside magnetospheric emission in future work \citep[see][and references therein]{DuncanTsang+2021}.
In the future, as models for EM emission in non-disrupting BHNSs mature, a compendium of signatures at different observing bands should be generated and used to place a range of model dependent magnetic field constraints.

\section{Conclusion}
\label{S:Conclusion}

We have presented the first multi-messenger constraints on neutron-star magnetic field strengths, providing a rare observational handle on the magnetic fields of neutron stars at these likely late stages of their lives, before merger with a black hole. The present observations require magnetic field strengths below $\sim$$10^{15}$~G, in the realm of magnetar field strengths. This rules out the theoretically possible, high-strength long-lived or rejuvenated fields at merger, and is consistent with the non-expectation of mergers of black holes and young magnetars. Future observations with present technology may enhance these constraints by adding a few similar multi-messenger BHNS constraints per year, with the expectation of a nearby ($\lesssim 100$ Mpc) event, which could constrain field strengths below $10^{13-14}$~G, in the next ten years. The limiting factor in achieving better constraints is the gamma-ray flux sensitivity -- more sensitive EM upper-limits would enhance these constraints with the square root of the flux sensitivity.

Observations with the power to constrain the neutron-star magnetic field at astrophysically interesting scales warrants further understanding of the BHNS magnetospheric physics. This will lead to more accurate predictions for the expected frequency-dependent flux and timing of radiation from these systems, which will contribute to searches for these signatures and to tighter constraints from non-detection. It will also hone predictions for multi-wavelength signatures of these events, which can be leveraged in addition to multi-messenger constraints as a more powerful probe of the BHNS magnetosphere and neutron star magnetic field evolution.

\acknowledgements
We thank Yuri Levin for useful discussions.
DJD received funding from the European Union's Horizon 2020
research and innovation programme under Marie Sklodowska-Curie grant agreement No. 101029157.
ZH acknowledges support by NASA grant NNX17AL82G and NSF grants 1715661 and AST-2006176. JL is supported in part by the Tow Foundation.
AVG acknowledges support by the Danish National Research Foundation (DNRF132).
J.S. is supported by the European Union's Horizon 2020 research and innovation programme under Marie Sklodowska-Curie grant agreement No. 844629 and through Villum Fonden grant No. 29466.

\bibliographystyle{apj} 
\bibliography{refs}

\begin{thebibliography}{}
\expandafter\ifx\csname natexlab\endcsname\relax\def\natexlab#1{#1}\fi

\bibitem[{{Abbott} {et~al.}(2020){Abbott}, {Abbott}, {Abraham}, {Acernese},
  {Ackley}, {Adams}, {Adhikari}, {Adya}, {Affeldt}, {Agathos}, {Agatsuma},
  {Aggarwal}, {Aguiar}, {Aich}, {Aiello}, {Ain}, {Ajith}, {Akcay}, {Allen},
  {Allocca}, {Altin}, {Amato}, {Anand}, {Ananyeva}, {Anderson}, {Anderson},
  {Angelova}, {Ansoldi}, {Antier}, {Appert}, {Arai}, {Araya}, {Areeda},
  {Ar{\`e}ne}, {Arnaud}, {Aronson}, {Arun}, {Asali}, {Ascenzi}, {Ashton},
  {Aston}, {Astone}, {Aubin}, {Aufmuth}, {AultONeal}, {Austin}, {Avendano},
  {Babak}, {Bacon}, {Badaracco}, {Bader}, {Bae}, {Baer}, {Baird}, {Baldaccini},
  {Ballardin}, {Ballmer}, {Bals}, {Balsamo}, {Baltus}, {Banagiri}, {Bankar},
  {Bankar}, {Barayoga}, {Barbieri}, {Barish}, {Barker}, {Barkett}, {Barneo},
  {Barone}, {Barr}, {Barsotti}, {Barsuglia}, {Barta}, {Bartlett}, {Bartos},
  {Bassiri}, {Basti}, {Bawaj}, {Bayley}, {Bazzan}, {B{\'e}csy}, {Bejger},
  {Belahcene}, {Bell}, {Beniwal}, {Benjamin}, {Benkel}, {Bentley}, {Bergamin},
  {Berger}, {Bergmann}, {Bernuzzi}, {Berry}, {Bersanetti}, {Bertolini},
  {Betzwieser}, {Bhandare}, {Bhandari}, {Bidler}, {Biggs}, {Bilenko},
  {Billingsley}, {Birney}, {Birnholtz}, {Biscans}, {Bischi}, {Biscoveanu},
  {Bisht}, {Bissenbayeva}, {Bitossi}, {Bizouard}, {Blackburn}, {Blackman},
  {Blair}, {Blair}, {Blair}, {Bobba}, {Bode}, {Boer}, {Boetzel}, {Bogaert},
  {Bondu}, {Bonilla}, {Bonnand}, {Booker}, {Boom}, {Bork}, {Boschi}, {Bose},
  {Bossilkov}, {Bosveld}, {Bouffanais}, {Bozzi}, {Bradaschia}, {Brady},
  {Bramley}, {Branchesi}, {Brau}, {Breschi}, {Briant}, {Briggs}, {Brighenti},
  {Brillet}, {Brinkmann}, {Brito}, {Brockill}, {Brooks}, {Brooks}, {Brown},
  {Brunett}, {Bruno}, {Bruntz}, {Buikema}, {Bulik}, {Bulten}, {Buonanno},
  {Buskulic}, {Byer}, {Cabero}, {Cadonati}, {Cagnoli}, {Cahillane}, {Bustillo},
  {Callaghan}, {Callister}, {Calloni}, {Camp}, {Canepa}, {Cannon}, {Cao},
  {Cao}, {Carapella}, {Carbognani}, {Caride}, {Carney}, {Carullo}, {Diaz},
  {Casentini}, {Casta{\~n}eda}, {Caudill}, {Cavagli{\`a}}, {Cavalier},
  {Cavalieri}, {Cella}, {Cerd{\'a}-Dur{\'a}n}, {Cesarini}, {Chaibi},
  {Chakravarti}, {Chan}, {Chan}, {Chao}, {Charlton}, {Chase},
  {Chassande-Mottin}, {Chatterjee}, {Chaturvedi}, {Chatziioannou}, {Chen},
  {Chen}, {Chen}, {Cheng}, {Cheong}, {Chia}, {Chiadini}, {Chierici},
  {Chincarini}, {Chiummo}, {Cho}, {Cho}, {Cho}, {Christensen}, {Chu}, {Chua},
  {Chung}, {Chung}, {Ciani}, {Ciecielag}, {Cie{\'s}lar}, {Ciobanu}, {Ciolfi},
  {Cipriano}, {Cirone}, {Clara}, {Clark}, {Clearwater}, {Clesse}, {Cleva},
  {Coccia}, {Cohadon}, {Cohen}, {Colleoni}, {Collette}, {Collins}, {Colpi},
  {Constancio}, {Conti}, {Cooper}, {Corban}, {Corbitt}, {Cordero-Carri{\'o}n},
  {Corezzi}, {Corley}, {Cornish}, {Corre}, {Corsi}, {Cortese}, {Costa},
  {Cotesta}, {Coughlin}, {Coughlin}, {Coulon}, {Countryman}, {Couvares},
  {Covas}, {Coward}, {Cowart}, {Coyne}, {Coyne}, {Creighton}, {Creighton},
  {Cripe}, {Croquette}, {Crowder}, {Cudell}, {Cullen}, {Cumming}, {Cummings},
  {Cunningham}, {Cuoco}, {Curylo}, {Canton}, {D{\'a}lya}, {Dana},
  {Daneshgaran-Bajastani}, {D'Angelo}, {Danilishin}, {D'Antonio}, {Danzmann},
  {Darsow-Fromm}, {Dasgupta}, {Datrier}, {Dattilo}, {Dave}, {Davier}, {Davies},
  {Davis}, {Daw}, {DeBra}, {Deenadayalan}, {Degallaix}, {De Laurentis},
  {Del{\'e}glise}, {Delfavero}, {De Lillo}, {Del Pozzo}, {DeMarchi},
  {D'Emilio}, {Demos}, {Dent}, {De Pietri}, {De Rosa}, {De Rossi}, {DeSalvo},
  {de Varona}, {Dhurandhar}, {D{\'\i}az}, {Diaz-Ortiz}, {Dietrich}, {Di Fiore},
  {Di Fronzo}, {Di Giorgio}, {Di Giovanni}, {Di Giovanni}, {Di Girolamo}, {Di
  Lieto}, {Ding}, {Di Pace}, {Di Palma}, {Di Renzo}, {Divakarla}, {Dmitriev},
  {Doctor}, {Donovan}, {Dooley}, {Doravari}, {Dorrington}, {Downes}, {Drago},
  {Driggers}, {Du}, {Ducoin}, {Dupej}, {Durante}, {D'Urso}, {Dwyer}, {Easter},
  {Eddolls}, {Edelman}, {Edo}, {Edy}, {Effler}, {Ehrens}, {Eichholz},
  {Eikenberry}, {Eisenmann}, {Eisenstein}, {Ejlli}, {Errico}, {Essick},
  {Estelles}, {Estevez}, {Etienne}, {Etzel}, {Evans}, {Evans}, {Ewing},
  {Fafone}, {Fairhurst}, {Fan}, {Farinon}, {Farr}, {Farr}, {Fauchon-Jones},
  {Favata}, {Fays}, {Fazio}, {Feicht}, {Fejer}, {Feng}, {Fenyvesi}, {Ferguson},
  {Fernandez-Galiana}, {Ferrante}, {Ferreira}, {Ferreira}, {Fidecaro}, {Fiori},
  {Fiorucci}, {Fishbach}, {Fisher}, {Fittipaldi}, {Fitz-Axen}, {Fiumara},
  {Flaminio}, {Floden}, {Flynn}, {Fong}, {Font}, {Forsyth}, {Fournier},
  {Frasca}, {Frasconi}, {Frei}, {Freise}, {Frey}, {Frey}, {Fritschel},
  {Frolov}, {Fronz{\`e}}, {Fulda}, {Fyffe}, {Gabbard}, {Gadre}, {Gaebel},
  {Gair}, {Galaudage}, {Ganapathy}, {Ganguly}, {Gaonkar},
  {Garc{\'\i}a-Quir{\'o}s}, {Garufi}, {Gateley}, {Gaudio}, {Gayathri}, {Gemme},
  {Genin}, {Gennai}, {George}, {George}, {Gergely}, {Ghonge}, {Ghosh}, {Ghosh},
  {Ghosh}, {Giacomazzo}, {Giaime}, {Giardina}, {Gibson}, {Gier}, {Gill},
  {Glanzer}, {Gniesmer}, {Godwin}, {Goetz}, {Goetz}, {Gohlke}, {Goncharov},
  {Gonz{\'a}lez}, {Gopakumar}, {Gossan}, {Gosselin}, {Gouaty}, {Grace},
  {Grado}, {Granata}, {Grant}, {Gras}, {Grassia}, {Gray}, {Gray}, {Greco},
  {Green}, {Green}, {Gretarsson}, {Griggs}, {Grignani}, {Grimaldi}, {Grimm},
  {Grote}, {Grunewald}, {Gruning}, {Guidi}, {Guimaraes}, {Guix{\'e}}, {Gulati},
  {Guo}, {Gupta}, {Gupta}, {Gupta}, {Gustafson}, {Gustafson}, {Haegel},
  {Halim}, {Hall}, {Hamilton}, {Hammond}, {Haney}, {Hanke}, {Hanks}, {Hanna},
  {Hannam}, {Hannuksela}, {Hansen}, {Hanson}, {Harder}, {Hardwick}, {Haris},
  {Harms}, {Harry}, {Harry}, {Hasskew}, {Haster}, {Haughian}, {Hayes}, {Healy},
  {Heidmann}, {Heintze}, {Heinze}, {Heitmann}, {Hellman}, {Hello}, {Hemming},
  {Hendry}, {Heng}, {Hennes}, {Hennig}, {Heurs}, {Hild}, {Hinderer}, {Hoback},
  {Hochheim}, {Hofgard}, {Hofman}, {Holgado}, {Holland}, {Holt}, {Holz},
  {Hopkins}, {Horst}, {Hough}, {Howell}, {Hoy}, {Huang}, {H{\"u}bner},
  {Huerta}, {Huet}, {Hughey}, {Hui}, {Husa}, {Huttner}, {Huxford},
  {Huynh-Dinh}, {Idzkowski}, {Iess}, {Inchauspe}, {Ingram}, {Intini}, {Isac},
  {Isi}, {Iyer}, {Jacqmin}, {Jadhav}, {Jadhav}, {James}, {Jani}, {Janthalur},
  {Jaranowski}, {Jariwala}, {Jaume}, {Jenkins}, {Jiang}, {Johns},
  {Johnson-McDaniel}, {Jones}, {Jones}, {Jones}, {Jones}, {Jones}, {Jonker},
  {Ju}, {Junker}, {Kalaghatgi}, {Kalogera}, {Kamai}, {Kandhasamy}, {Kang},
  {Kanner}, {Kapadia}, {Karki}, {Kashyap}, {Kasprzack}, {Kastaun},
  {Katsanevas}, {Katsavounidis}, {Katzman}, {Kaufer}, {Kawabe},
  {K{\'e}f{\'e}lian}, {Keitel}, {Keivani}, {Kennedy}, {Key}, {Khadka},
  {Khalili}, {Khan}, {Khan}, {Khan}, {Khazanov}, {Khetan}, {Khursheed},
  {Kijbunchoo}, {Kim}, {Kim}, {Kim}, {Kim}, {Kim}, {Kim}, {Kim}, {Kimball},
  {King}, {Kinley-Hanlon}, {Kirchhoff}, {Kissel}, {Kleybolte}, {Klimenko},
  {Knowles}, {Knyazev}, {Koch}, {Koehlenbeck}, {Koekoek}, {Koley},
  {Kondrashov}, {Kontos}, {Koper}, {Korobko}, {Korth}, {Kovalam}, {Kozak},
  {Kringel}, {Krishnendu}, {Kr{\'o}lak}, {Krupinski}, {Kuehn}, {Kumar},
  {Kumar}, {Kumar}, {Kumar}, {Kumar}, {Kuo}, {Kutynia}, {Lackey}, {Laghi},
  {Lalande}, {Lam}, {Lamberts}, {Landry}, {Landry}, {Lane}, {Lang}, {Lange},
  {Lantz}, {Lanza}, {La Rosa}, {Lartaux-Vollard}, {Lasky}, {Laxen},
  {Lazzarini}, {Lazzaro}, {Leaci}, {Leavey}, {Lecoeuche}, {Lee}, {Lee}, {Lee},
  {Lee}, {Lee}, {Lehmann}, {Leroy}, {Letendre}, {Levin}, {Li}, {Li}, {li},
  {Li}, {Li}, {Linde}, {Linker}, {Linley}, {Littenberg}, {Liu}, {Liu},
  {Llorens-Monteagudo}, {Lo}, {Lockwood}, {London}, {Longo}, {Lorenzini},
  {Loriette}, {Lormand}, {Losurdo}, {Lough}, {Lousto}, {Lovelace}, {L{\"u}ck},
  {Lumaca}, {Lundgren}, {Ma}, {Macas}, {Macfoy}, {MacInnis}, {Macleod},
  {MacMillan}, {Macquet}, {Hernandez}, {Maga{\~n}a-Sandoval}, {Magee},
  {Majorana}, {Maksimovic}, {Malik}, {Man}, {Mandic}, {Mangano}, {Mansell},
  {Manske}, {Mantovani}, {Mapelli}, {Marchesoni}, {Marion}, {M{\'a}rka},
  {M{\'a}rka}, {Markakis}, {Markosyan}, {Markowitz}, {Maros}, {Marquina},
  {Marsat}, {Martelli}, {Martin}, {Martin}, {Martinez}, {Martynov},
  {Masalehdan}, {Mason}, {Massera}, {Masserot}, {Massinger}, {Masso-Reid},
  {Mastrogiovanni}, {Matas}, {Matichard}, {Mavalvala}, {Maynard}, {McCann},
  {McCarthy}, {McClelland}, {McCormick}, {McCuller}, {McGuire}, {McIsaac},
  {McIver}, {McManus}, {McRae}, {McWilliams}, {Meacher}, {Meadors}, {Mehmet},
  {Mehta}, {Villa}, {Melatos}, {Mendell}, {Mercer}, {Mereni}, {Merfeld},
  {Merilh}, {Merritt}, {Merzougui}, {Meshkov}, {Messenger}, {Messick},
  {Metzdorff}, {Meyers}, {Meylahn}, {Mhaske}, {Miani}, {Miao}, {Michaloliakos},
  {Michel}, {Middleton}, {Milano}, {Miller}, {Millhouse}, {Mills}, {Milotti},
  {Milovich-Goff}, {Minazzoli}, {Minenkov}, {Mishkin}, {Mishra}, {Mistry},
  {Mitra}, {Mitrofanov}, {Mitselmakher}, {Mittleman}, {Mo}, {Mogushi},
  {Mohapatra}, {Mohite}, {Molina-Ruiz}, {Mondin}, {Montani}, {Moore}, {Moraru},
  {Morawski}, {Moreno}, {Morisaki}, {Mours}, {Mow-Lowry}, {Mozzon},
  {Muciaccia}, {Mukherjee}, {Mukherjee}, {Mukherjee}, {Mukherjee}, {Mukund},
  {Mullavey}, {Munch}, {Mu{\~n}iz}, {Murray}, {Nagar}, {Nardecchia},
  {Naticchioni}, {Nayak}, {Neil}, {Neilson}, {Nelemans}, {Nelson}, {Nery},
  {Neunzert}, {Ng}, {Ng}, {Nguyen}, {Nguyen}, {Nichols}, {Nichols}, {Nissanke},
  {Nocera}, {Noh}, {North}, {Nothard}, {Nuttall}, {Oberling}, {O'Brien},
  {Oganesyan}, {Ogin}, {Oh}, {Oh}, {Ohme}, {Ohta}, {Okada}, {Oliver},
  {Olivetto}, {Oppermann}, {Oram}, {O'Reilly}, {Ormiston}, {Ortega},
  {O'Shaughnessy}, {Ossokine}, {Osthelder}, {Ottaway}, {Overmier}, {Owen},
  {Pace}, {Pagano}, {Page}, {Pagliaroli}, {Pai}, {Pai}, {Palamos}, {Palashov},
  {Palomba}, {Pan}, {Panda}, {Pang}, {Pankow}, {Pannarale}, {Pant}, {Paoletti},
  {Paoli}, {Parida}, {Parker}, {Pascucci}, {Pasqualetti}, {Passaquieti},
  {Passuello}, {Patricelli}, {Payne}, {Pearlstone}, {Pechsiri}, {Pedersen},
  {Pedraza}, {Pele}, {Penn}, {Perego}, {Perez}, {P{\'e}rigois}, {Perreca},
  {Perri{\`e}s}, {Petermann}, {Pfeiffer}, {Phelps}, {Phukon}, {Piccinni},
  {Pichot}, {Piendibene}, {Piergiovanni}, {Pierro}, {Pillant}, {Pinard},
  {Pinto}, {Piotrzkowski}, {Pirello}, {Pitkin}, {Plastino}, {Poggiani}, {Pong},
  {Ponrathnam}, {Popolizio}, {Porter}, {Powell}, {Prajapati}, {Prasai},
  {Prasanna}, {Pratten}, {Prestegard}, {Principe}, {Prodi}, {Prokhorov},
  {Punturo}, {Puppo}, {P{\"u}rrer}, {Qi}, {Quetschke}, {Quinonez}, {Raab},
  {Raaijmakers}, {Radkins}, {Radulesco}, {Raffai}, {Rafferty}, {Raja}, {Rajan},
  {Rajbhandari}, {Rakhmanov}, {Ramirez}, {Ramos-Buades}, {Rana}, {Rao},
  {Rapagnani}, {Raymond}, {Razzano}, {Read}, {Regimbau}, {Rei}, {Reid},
  {Reitze}, {Rettegno}, {Ricci}, {Richardson}, {Richardson}, {Ricker},
  {Riemenschneider}, {Riles}, {Rizzo}, {Robertson}, {Robinet}, {Rocchi},
  {Rodriguez-Soto}, {Rolland}, {Rollins}, {Roma}, {Romanelli}, {Romano},
  {Romel}, {Romero-Shaw}, {Romie}, {Rose}, {Rose}, {Rose}, {Rosi{\'n}ska},
  {Rosofsky}, {Ross}, {Rowan}, {Rowlinson}, {Roy}, {Roy}, {Roy}, {Ruggi},
  {Rutins}, {Ryan}, {Sachdev}, {Sadecki}, {Sakellariadou}, {Salafia},
  {Salconi}, {Saleem}, {Salemi}, {Samajdar}, {Sanchez}, {Sanchez},
  {Sanchis-Gual}, {Sanders}, {Santiago}, {Santos}, {Sarin}, {Sassolas},
  {Sathyaprakash}, {Sauter}, {Savage}, {Savant}, {Sawant}, {Sayah}, {Schaetzl},
  {Schale}, {Scheel}, {Scheuer}, {Schmidt}, {Schnabel}, {Schofield},
  {Sch{\"o}nbeck}, {Schreiber}, {Schulte}, {Schutz}, {Schwarm}, {Schwartz},
  {Scott}, {Scott}, {Seidel}, {Sellers}, {Sengupta}, {Sennett}, {Sentenac},
  {Sequino}, {Sergeev}, {Setyawati}, {Shaddock}, {Shaffer}, {Shahriar},
  {Sharma}, {Sharma}, {Shawhan}, {Shen}, {Shikauchi}, {Shink}, {Shoemaker},
  {Shoemaker}, {Shukla}, {ShyamSundar}, {Siellez}, {Sieniawska}, {Sigg},
  {Singer}, {Singh}, {Singh}, {Singha}, {Singhal}, {Sintes}, {Sipala},
  {Skliris}, {Slagmolen}, {Slaven-Blair}, {Smetana}, {Smith}, {Smith},
  {Somala}, {Son}, {Soni}, {Sorazu}, {Sordini}, {Sorrentino}, {Souradeep},
  {Sowell}, {Spencer}, {Spera}, {Srivastava}, {Srivastava}, {Staats},
  {Stachie}, {Standke}, {Steer}, {Steinhoff}, {Steinke}, {Steinlechner},
  {Steinlechner}, {Steinmeyer}, {Stevenson}, {Stocks}, {Stops}, {Stover},
  {Strain}, {Stratta}, {Strunk}, {Sturani}, {Stuver}, {Sudhagar}, {Sudhir},
  {Summerscales}, {Sun}, {Sunil}, {Sur}, {Suresh}, {Sutton}, {Swinkels},
  {Szczepa{\'n}czyk}, {Tacca}, {Tait}, {Talbot}, {Tanasijczuk}, {Tanner},
  {Tao}, {T{\'a}pai}, {Tapia}, {San Martin}, {Tasson}, {Taylor}, {Tenorio},
  {Terkowski}, {Thirugnanasambandam}, {Thomas}, {Thomas}, {Thompson},
  {Thondapu}, {Thorne}, {Thrane}, {Tinsman}, {Saravanan}, {Tiwari}, {Tiwari},
  {Tiwari}, {Toland}, {Tonelli}, {Tornasi}, {Torres-Forn{\'e}}, {Torrie},
  {Tosta e Melo}, {T{\"o}yr{\"a}}, {Trail}, {Travasso}, {Traylor}, {Tringali},
  {Tripathee}, {Trovato}, {Trudeau}, {Tsang}, {Tse}, {Tso}, {Tsukada}, {Tsuna},
  {Tsutsui}, {Turconi}, {Ubhi}, {Ueno}, {Ugolini}, {Unnikrishnan}, {Urban},
  {Usman}, {Utina}, {Vahlbruch}, {Vajente}, {Valdes}, {Valentini}, {van Bakel},
  {van Beuzekom}, {van den Brand}, {Van Den Broeck}, {Vander-Hyde}, {van der
  Schaaf}, {Van Heijningen}, {van Veggel}, {Vardaro}, {Varma}, {Vass},
  {Vas{\'u}th}, {Vecchio}, {Vedovato}, {Veitch}, {Veitch}, {Venkateswara},
  {Venugopalan}, {Verkindt}, {Veske}, {Vetrano}, {Vicer{\'e}}, {Viets},
  {Vinciguerra}, {Vine}, {Vinet}, {Vitale}, {Vivanco}, {Vo}, {Vocca},
  {Vorvick}, {Vyatchanin}, {Wade}, {Wade}, {Wade}, {Walet}, {Walker},
  {Wallace}, {Wallace}, {Walsh}, {Wang}, {Wang}, {Wang}, {Ward}, {Warden},
  {Warner}, {Was}, {Watchi}, {Weaver}, {Wei}, {Weinert}, {Weinstein}, {Weiss},
  {Wellmann}, {Wen}, {We{\ss}els}, {Westhouse}, {Wette}, {Whelan}, {Whiting},
  {Whittle}, {Wilken}, {Williams}, {Willis}, {Willke}, {Winkler}, {Wipf},
  {Wittel}, {Woan}, {Woehler}, {Wofford}, {Wong}, {Wright}, {Wu}, {Wysocki},
  {Xiao}, {Yamamoto}, {Yang}, {Yang}, {Yang}, {Yap}, {Yazback}, {Yeeles}, {Yu},
  {Yu}, {Yuen}, {Zadro{\.z}ny}, {Zadro{\.z}ny}, {Zanolin}, {Zelenova},
  {Zendri}, {Zevin}, {Zhang}, {Zhang}, {Zhang}, {Zhao}, {Zhao}, {Zhou}, {Zhou},
  {Zhu}, {Zimmerman}, {Zucker}, {Zweizig}, {LIGO Scientific Collaboration}, \&
  {Virgo Collaboration}}]{GW190814:2020}
{Abbott}, R., {Abbott}, T.~D., {Abraham}, S., {et~al.} 2020, \apjl, 896, L44

\bibitem[{{Abbott} {et~al.}(2021){Abbott}, {Abbott}, {Abraham}, {Acernese},
  {Ackley}, {Adams}, {Adams}, {Adhikari}, {Adya}, {Affeldt}, {Agarwal},
  {Agathos}, {Agatsuma}, {Aggarwal}, {Aguiar}, {Aiello}, {Ain}, {Ajith},
  {Akutsu}, {Aleman}, {Allen}, {Allocca}, {Altin}, {Amato}, {Anand},
  {Ananyeva}, {Anderson}, {Anderson}, {Ando}, {Angelova}, {Ansoldi}, {Antelis},
  {Antier}, {Appert}, {Arai}, {Arai}, {Arai}, {Araki}, {Araya}, {Araya},
  {Areeda}, {Ar{\`e}ne}, {Aritomi}, {Arnaud}, {Aronson}, {Arun}, {Asada},
  {Asali}, {Ashton}, {Aso}, {Aston}, {Astone}, {Aubin}, {Aufmuth}, {Aultoneal},
  {Austin}, {Babak}, {Badaracco}, {Bader}, {Bae}, {Bae}, {Baer}, {Bagnasco},
  {Bai}, {Baiotti}, {Baird}, {Bajpai}, {Ball}, {Ballardin}, {Ballmer}, {Bals},
  {Balsamo}, {Baltus}, {Banagiri}, {Bankar}, {Bankar}, {Barayoga}, {Barbieri},
  {Barish}, {Barker}, {Barneo}, {Barone}, {Barr}, {Barsotti}, {Barsuglia},
  {Barta}, {Bartlett}, {Barton}, {Bartos}, {Bassiri}, {Basti}, {Bawaj},
  {Bayley}, {Baylor}, {Bazzan}, {B{\'e}csy}, {Bedakihale}, {Bejger},
  {Belahcene}, {Benedetto}, {Beniwal}, {Benjamin}, {Benkel}, {Bennett},
  {Bentley}, {Benyaala}, {Bergamin}, {Berger}, {Bernuzzi}, {Berry},
  {Bersanetti}, {Bertolini}, {Betzwieser}, {Bhandare}, {Bhandari},
  {Bhattacharjee}, {Bhaumik}, {Bidler}, {Bilenko}, {Billingsley}, {Birney},
  {Birnholtz}, {Biscans}, {Bischi}, {Biscoveanu}, {Bisht}, {Biswas}, {Bitossi},
  {Bizouard}, {Blackburn}, {Blackman}, {Blair}, {Blair}, {Blair}, {Bobba},
  {Bode}, {Boer}, {Bogaert}, {Boldrini}, {Bondu}, {Bonilla}, {Bonnand},
  {Booker}, {Boom}, {Bork}, {Boschi}, {Bose}, {Bose}, {Bossilkov}, {Boudart},
  {Bouffanais}, {Bozzi}, {Bradaschia}, {Brady}, {Bramley}, {Branch},
  {Branchesi}, {Brau}, {Breschi}, {Briant}, {Briggs}, {Brillet}, {Brinkmann},
  {Brockill}, {Brooks}, {Brooks}, {Brown}, {Brunett}, {Bruno}, {Bruntz},
  {Bryant}, {Buikema}, {Bulik}, {Bulten}, {Buonanno}, {Buscicchio}, {Buskulic},
  {Byer}, {Cadonati}, {Caesar}, {Cagnoli}, {Cahillane}, {Cain}, {Calder{\'o}n
  Bustillo}, {Callaghan}, {Callister}, {Calloni}, {Camp}, {Canepa},
  {Cannavacciuolo}, {Cannon}, {Cao}, {Cao}, {Cao}, {Capocasa}, {Capote},
  {Carapella}, {Carbognani}, {Carlin}, {Carney}, {Carpinelli}, {Carullo},
  {Carver}, {Casanueva Diaz}, {Casentini}, {Castaldi}, {Caudill},
  {Cavagli{\`a}}, {Cavalier}, {Cavalieri}, {Cella}, {Cerd{\'a}-Dur{\'a}n},
  {Cesarini}, {Chaibi}, {Chakravarti}, {Champion}, {Chan}, {Chan}, {Chan},
  {Chan}, {Chandra}, {Chanial}, {Chao}, {Charlton}, {Chase},
  {Chassande-Mottin}, {Chatterjee}, {Chaturvedi}, {Chatziioannou}, {Chen},
  {Chen}, {Chen}, {Chen}, {Chen}, {Chen}, {Chen}, {Chen}, {Chen}, {Cheng},
  {Cheong}, {Cheung}, {Chia}, {Chiadini}, {Chiang}, {Chierici}, {Chincarini},
  {Chiofalo}, {Chiummo}, {Cho}, {Cho}, {Choate}, {Choudhary}, {Choudhary},
  {Christensen}, {Chu}, {Chu}, {Chu}, {Chua}, {Chung}, {Ciani}, {Ciecielag},
  {Cie{\'s}lar}, {Cifaldi}, {Ciobanu}, {Ciolfi}, {Cipriano}, {Cirone}, {Clara},
  {Clark}, {Clark}, {Clarke}, {Clearwater}, {Clesse}, {Cleva}, {Coccia},
  {Cohadon}, {Cohen}, {Cohen}, {Colleoni}, {Collette}, {Colpi}, {Compton},
  {Constancio}, {Conti}, {Cooper}, {Corban}, {Corbitt}, {Cordero-Carri{\'o}n},
  {Corezzi}, {Corley}, {Cornish}, {Corre}, {Corsi}, {Cortese}, {Costa},
  {Cotesta}, {Coughlin}, {Coughlin}, {Coulon}, {Countryman}, {Cousins},
  {Couvares}, {Covas}, {Coward}, {Cowart}, {Coyne}, {Coyne}, {Creighton},
  {Creighton}, {Criswell}, {Croquette}, {Crowder}, {Cudell}, {Cullen},
  {Cumming}, {Cummings}, {Cuoco}, {Cury{\l}o}, {Dal Canton}, {D{\'a}lya},
  {Dana}, {Daneshgaranbajastani}, {D'Angelo}, {Danilishin}, {D'Antonio},
  {Danzmann}, {Darsow-Fromm}, {Dasgupta}, {Datrier}, {Dattilo}, {Dave},
  {Davier}, {Davies}, {Davis}, {Daw}, {Dean}, {Debra}, {Deenadayalan},
  {Degallaix}, {de Laurentis}, {Del{\'e}glise}, {Del Favero}, {de Lillo}, {de
  Lillo}, {Del Pozzo}, {Demarchi}, {de Matteis}, {D'Emilio}, {Demos}, {Dent},
  {Depasse}, {de Pietri}, {De Rosa}, {de Rossi}, {Desalvo}, {de Simone},
  {Dhurandhar}, {D{\'\i}az}, {Diaz-Ortiz}, {Didio}, {Dietrich}, {di Fiore}, {di
  Fronzo}, {di Giorgio}, {di Giovanni}, {di Girolamo}, {di Lieto}, {Ding}, {di
  Pace}, {di Palma}, {di Renzo}, {Divakarla}, {Dmitriev}, {Doctor},
  {D'Onofrio}, {Donovan}, {Dooley}, {Doravari}, {Dorrington}, {Drago},
  {Driggers}, {Drori}, {Du}, {Ducoin}, {Dupej}, {Durante}, {D'Urso}, {Duverne},
  {Dwyer}, {Easter}, {Ebersold}, {Eddolls}, {Edelman}, {Edo}, {Edy}, {Effler},
  {Eguchi}, {Eichholz}, {Eikenberry}, {Eisenmann}, {Eisenstein}, {Ejlli},
  {Enomoto}, {Errico}, {Essick}, {Estell{\'e}s}, {Estevez}, {Etienne}, {Etzel},
  {Evans}, {Evans}, {Ewing}, {Fafone}, {Fair}, {Fairhurst}, {Fan}, {Farah},
  {Farinon}, {Farr}, {Farr}, {Farrow}, {Fauchon-Jones}, {Favata}, {Fays},
  {Fazio}, {Feicht}, {Fejer}, {Feng}, {Fenyvesi}, {Ferguson},
  {Fernandez-Galiana}, {Ferrante}, {Ferreira}, {Fidecaro}, {Figura}, {Fiori},
  {Fishbach}, {Fisher}, {Fittipaldi}, {Fiumara}, {Flaminio}, {Floden}, {Flynn},
  {Fong}, {Font}, {Fornal}, {Forsyth}, {Franke}, {Frasca}, {Frasconi},
  {Frederick}, {Frei}, {Freise}, {Frey}, {Fritschel}, {Frolov}, {Fronz{\'e}},
  {Fujii}, {Fujikawa}, {Fukunaga}, {Fukushima}, {Fulda}, {Fyffe}, {Gabbard},
  {Gadre}, {Gaebel}, {Gair}, {Gais}, {Galaudage}, {Gamba}, {Ganapathy},
  {Ganguly}, {Gao}, {Gaonkar}, {Garaventa}, {Garc{\'\i}a-N{\'u}{\~n}ez},
  {Garc{\'\i}a-Quir{\'o}s}, {Garufi}, {Gateley}, {Gaudio}, {Gayathri}, {Ge},
  {Gemme}, {Gennai}, {George}, {Gergely}, {Gewecke}, {Ghonge}, {Ghosh},
  {Ghosh}, {Ghosh}, {Ghosh}, {Ghosh}, {Giacomazzo}, {Giacoppo}, {Giaime},
  {Giardina}, {Gibson}, {Gier}, {Giesler}, {Giri}, {Gissi}, {Glanzer},
  {Gleckl}, {Godwin}, {Goetz}, {Goetz}, {Gohlke}, {Goncharov}, {Gonz{\'a}lez},
  {Gopakumar}, {Gosselin}, {Gouaty}, {Grace}, {Grado}, {Granata}, {Granata},
  {Grant}, {Gras}, {Grassia}, {Gray}, {Gray}, {Greco}, {Green}, {Green},
  {Gretarsson}, {Gretarsson}, {Griffith}, {Griffiths}, {Griggs}, {Grignani},
  {Grimaldi}, {Grimes}, {Grimm}, {Grote}, {Grunewald}, {Gruning}, {Guerrero},
  {Guidi}, {Guimaraes}, {Guix{\'e}}, {Gulati}, {Guo}, {Guo}, {Gupta}, {Gupta},
  {Gupta}, {Gustafson}, {Gustafson}, {Guzman}, {Ha}, {Haegel}, {Hagiwara},
  {Haino}, {Halim}, {Hall}, {Hamilton}, {Hammond}, {Han}, {Haney}, {Hanks},
  {Hanna}, {Hannam}, {Hannuksela}, {Hansen}, {Hansen}, {Hanson}, {Harder},
  {Hardwick}, {Haris}, {Harms}, {Harry}, {Harry}, {Hartwig}, {Hasegawa},
  {Haskell}, {Hasskew}, {Haster}, {Hattori}, {Haughian}, {Hayakawa}, {Hayama},
  {Hayes}, {Healy}, {Heidmann}, {Heintze}, {Heinze}, {Heinzel}, {Heitmann},
  {Hellman}, {Hello}, {Helmling-Cornell}, {Hemming}, {Hendry}, {Heng},
  {Hennes}, {Hennig}, {Hennig}, {Hernandez Vivanco}, {Heurs}, {Hild}, {Hill},
  {Himemoto}, {Hinderer}, {Hines}, {Hiranuma}, {Hirata}, {Hirose}, {Ho},
  {Hochheim}, {Hofman}, {Hohmann}, {Holgado}, {Holland}, {Hollows}, {Holmes},
  {Holt}, {Holz}, {Hong}, {Hopkins}, {Hough}, {Howell}, {Hoy}, {Hoyland},
  {Hreibi}, {Hsieh}, {Hsu}, {Huang}, {Huang}, {Huang}, {Huang}, {Huang},
  {Huang}, {H{\"u}bner}, {Huddart}, {Huerta}, {Hughey}, {Hui}, {Hui}, {Husa},
  {Huttner}, {Huxford}, {Huynh-Dinh}, {Ide}, {Idzkowski}, {Iess}, {Ikenoue},
  {Imam}, {Inayoshi}, {Inchauspe}, {Ingram}, {Inoue}, {Intini}, {Ioka}, {Isi},
  {Isleif}, {Ito}, {Itoh}, {Iyer}, {Izumi}, {Jaberianhamedan}, {Jacqmin},
  {Jadhav}, {Jadhav}, {James}, {Jan}, {Jani}, {Janssens}, {Janthalur},
  {Jaranowski}, {Jariwala}, {Jaume}, {Jenkins}, {Jeon}, {Jeunon}, {Jia},
  {Jiang}, {Jin}, {Johns}, {Jones}, {Jones}, {Jones}, {Jones}, {Jones},
  {Jonker}, {Ju}, {Jung}, {Jung}, {Junker}, {Kaihotsu}, {Kajita}, {Kakizaki},
  {Kalaghatgi}, {Kalogera}, {Kamai}, {Kamiizumi}, {Kanda}, {Kandhasamy},
  {Kang}, {Kanner}, {Kao}, {Kapadia}, {Kapasi}, {Karat}, {Karathanasis},
  {Karki}, {Kashyap}, {Kasprzack}, {Kastaun}, {Katsanevas}, {Katsavounidis},
  {Katzman}, {Kaur}, {Kawabe}, {Kawaguchi}, {Kawai}, {Kawasaki},
  {K{\'e}f{\'e}lian}, {Keitel}, {Key}, {Khadka}, {Khalili}, {Khan}, {Khan},
  {Khazanov}, {Khetan}, {Khursheed}, {Kijbunchoo}, {Kim}, {Kim}, {Kim}, {Kim},
  {Kim}, {Kim}, {Kimball}, {Kimura}, {King}, {Kinley-Hanlon}, {Kirchhoff},
  {Kissel}, {Kita}, {Kitazawa}, {Kleybolte}, {Klimenko}, {Knee}, {Knowles},
  {Knyazev}, {Koch}, {Koekoek}, {Kojima}, {Kokeyama}, {Koley}, {Kolitsidou},
  {Kolstein}, {Komori}, {Kondrashov}, {Kong}, {Kontos}, {Koper}, {Korobko},
  {Kotake}, {Kovalam}, {Kozak}, {Kozakai}, {Kozu}, {Kringel}, {Krishnendu},
  {Kr{\'o}lak}, {Kuehn}, {Kuei}, {Kumar}, {Kumar}, {Kumar}, {Kumar}, {Kume},
  {Kuns}, {Kuo}, {Kuo}, {Kuromiya}, {Kuroyanagi}, {Kusayanagi}, {Kwak},
  {Kwang}, {Laghi}, {Lalande}, {Lam}, {Lamberts}, {Landry}, {Landry}, {Lane},
  {Lang}, {Lange}, {Lantz}, {La Rosa}, {Lartaux-Vollard}, {Lasky}, {Laxen},
  {Lazzarini}, {Lazzaro}, {Leaci}, {Leavey}, {Lecoeuche}, {Lee}, {Lee}, {Lee},
  {Lee}, {Lee}, {Lee}, {Lehmann}, {Lema{\^\i}tre}, {Leon}, {Leonardi}, {Leroy},
  {Letendre}, {Levin}, {Leviton}, {Li}, {Li}, {Li}, {Li}, {Li}, {Li}, {Lin},
  {Lin}, {Lin}, {Lin}, {Lin}, {Linde}, {Linker}, {Linley}, {Littenberg}, {Liu},
  {Liu}, {Liu}, {Liu}, {Llorens-Monteagudo}, {Lo}, {Lockwood}, {Lollie},
  {London}, {Longo}, {Lopez}, {Lorenzini}, {Loriette}, {Lormand}, {Losurdo},
  {Lough}, {Lousto}, {Lovelace}, {L{\"u}ck}, {Lumaca}, {Lundgren}, {Luo},
  {Macas}, {Macinnis}, {MacLeod}, {MacMillan}, {Macquet}, {Maga{\~n}a
  Hernandez}, {Maga{\~n}a-Sandoval}, {Magazz{\`u}}, {Magee}, {Maggiore},
  {Majorana}, {Makarem}, {Maksimovic}, {Maliakal}, {Malik}, {Man}, {Mandic},
  {Mangano}, {Mango}, {Mansell}, {Manske}, {Mantovani}, {Mapelli},
  {Marchesoni}, {Marchio}, {Marion}, {Mark}, {M{\'a}rka}, {M{\'a}rka},
  {Markakis}, {Markosyan}, {Markowitz}, {Maros}, {Marquina}, {Marsat},
  {Martelli}, {Martin}, {Martin}, {Martinez}, {Martinez}, {Martinovic},
  {Martynov}, {Marx}, {Masalehdan}, {Mason}, {Massera}, {Masserot},
  {Massinger}, {Masso-Reid}, {Mastrogiovanni}, {Matas}, {Mateu-Lucena},
  {Matichard}, {Matiushechkina}, {Mavalvala}, {McCann}, {McCarthy},
  {McClelland}, {McClincy}, {McCormick}, {McCuller}, {McGhee}, {McGuire},
  {McIsaac}, {McIver}, {McManus}, {McRae}, {McWilliams}, {Meacher}, {Mehmet},
  {Mehta}, {Melatos}, {Melchor}, {Mendell}, {Menendez-Vazquez}, {Menoni},
  {Mercer}, {Mereni}, {Merfeld}, {Merilh}, {Merritt}, {Merzougui}, {Meshkov},
  {Messenger}, {Messick}, {Meyers}, {Meylahn}, {Mhaske}, {Miani}, {Miao},
  {Michaloliakos}, {Michel}, {Michimura}, {Middleton}, {Milano}, {Miller},
  {Millhouse}, {Mills}, {Milotti}, {Milovich-Goff}, {Minazzoli}, {Minenkov},
  {Mio}, {Mir}, {Mishkin}, {Mishra}, {Mishra}, {Mistry}, {Mitra}, {Mitrofanov},
  {Mitselmakher}, {Mittleman}, {Miyakawa}, {Miyamoto}, {Miyazaki}, {Miyo},
  {Miyoki}, {Mo}, {Mogushi}, {Mohapatra}, {Mohite}, {Molina}, {Molina-Ruiz},
  {Mondin}, {Montani}, {Moore}, {Moraru}, {Morawski}, {More}, {Moreno},
  {Moreno}, {Mori}, {Morisaki}, {Moriwaki}, {Mours}, {Mow-Lowry}, {Mozzon},
  {Muciaccia}, {Mukherjee}, {Mukherjee}, {Mukherjee}, {Mukherjee}, {Mukund},
  {Mullavey}, {Munch}, {Mu{\~n}iz}, {Murray}, {Musenich}, {Nadji}, {Nagano},
  {Nagano}, {Nagar}, {Nakamura}, {Nakano}, {Nakano}, {Nakashima}, {Nakayama},
  {Nardecchia}, {Narikawa}, {Naticchioni}, {Nayak}, {Nayak}, {Negishi}, {Neil},
  {Neilson}, {Nelemans}, {Nelson}, {Nery}, {Neunzert}, {Ng}, {Ng}, {Nguyen},
  {Nguyen}, {Nguyen}, {Nguyen Quynh}, {Ni}, {Nichols}, {Nishizawa}, {Nissanke},
  {Nocera}, {Noh}, {Norman}, {North}, {Nozaki}, {Nuttall}, {Oberling},
  {O'Brien}, {Obuchi}, {O'Dell}, {Ogaki}, {Oganesyan}, {Oh}, {Oh}, {Oh},
  {Ohashi}, {Ohishi}, {Ohkawa}, {Ohme}, {Ohta}, {Okada}, {Okutani}, {Okutomi},
  {Olivetto}, {Oohara}, {Ooi}, {Oram}, {O'Reilly}, {Ormiston}, {Ormsby},
  {Ortega}, {O'Shaughnessy}, {O'Shea}, {Oshino}, {Ossokine}, {Osthelder},
  {Otabe}, {Ottaway}, {Overmier}, {Pace}, {Pagano}, {Page}, {Pagliaroli},
  {Pai}, {Pai}, {Palamos}, {Palashov}, {Palomba}, {Pan}, {Panda}, {Pang},
  {Pang}, {Pankow}, {Pannarale}, {Pant}, {Paoletti}, {Paoli}, {Paolone},
  {Parisi}, {Park}, {Parker}, {Pascucci}, {Pasqualetti}, {Passaquieti},
  {Passuello}, {Patel}, {Patricelli}, {Payne}, {Pechsiri}, {Pedraza},
  {Pegoraro}, {Pele}, {Pe{\~n}a Arellano}, {Penn}, {Perego}, {Pereira},
  {Pereira}, {Perez}, {P{\'e}rigois}, {Perreca}, {Perri{\`e}s}, {Petermann},
  {Petterson}, {Pfeiffer}, {Pham}, {Phukon}, {Piccinni}, {Pichot},
  {Piendibene}, {Piergiovanni}, {Pierini}, {Pierro}, {Pillant}, {Pilo},
  {Pinard}, {Pinto}, {Piotrzkowski}, {Piotrzkowski}, {Pirello}, {Pitkin},
  {Placidi}, {Plastino}, {Pluchar}, {Poggiani}, {Polini}, {Pong}, {Ponrathnam},
  {Popolizio}, {Porter}, {Powell}, {Pracchia}, {Pradier}, {Prajapati},
  {Prasai}, {Prasanna}, {Pratten}, {Prestegard}, {Principe}, {Prodi},
  {Prokhorov}, {Prosposito}, {Prudenzi}, {Puecher}, {Punturo}, {Puosi},
  {Puppo}, {P{\"u}rrer}, {Qi}, {Quetschke}, {Quinonez}, {Quitzow-James},
  {Raab}, {Raaijmakers}, {Radkins}, {Radulesco}, {Raffai}, {Rail}, {Raja},
  {Rajan}, {Ramirez}, {Ramirez}, {Ramos-Buades}, {Rana}, {Rapagnani}, {Rapol},
  {Ratto}, {Ray}, {Raymond}, {Raza}, {Razzano}, {Read}, {Rees}, {Regimbau},
  {Rei}, {Reid}, {Reitze}, {Relton}, {Rettegno}, {Ricci}, {Richardson},
  {Richardson}, {Richardson}, {Ricker}, {Riemenschneider}, {Riles}, {Rizzo},
  {Robertson}, {Robie}, {Robinet}, {Rocchi}, {Rocha}, {Rodriguez},
  {Rodriguez-Soto}, {Rolland}, {Rollins}, {Roma}, {Romanelli}, {Romano},
  {Romel}, {Romero}, {Romero-Shaw}, {Romie}, {Rose}, {Rosi{\'n}ska},
  {Rosofsky}, {Ross}, {Rowan}, {Rowlinson}, {Roy}, {Roy}, {Rozza}, {Ruggi},
  {Ryan}, {Sachdev}, {Sadecki}, {Sadiq}, {Sago}, {Saito}, {Saito}, {Sakai},
  {Sakai}, {Sakellariadou}, {Sakuno}, {Salafia}, {Salconi}, {Saleem}, {Salemi},
  {Samajdar}, {Sanchez}, {Sanchez}, {Sanchez}, {Sanchis-Gual}, {Sanders},
  {Sanuy}, {Saravanan}, {Sarin}, {Sassolas}, {Satari}, {Sathyaprakash}, {Sato},
  {Sato}, {Sauter}, {Savage}, {Savant}, {Sawada}, {Sawant}, {Sawant}, {Sayah},
  {Schaetzl}, {Scheel}, {Scheuer}, {Schindler-Tyka}, {Schmidt}, {Schnabel},
  {Schneewind}, {Schofield}, {Sch{\"o}nbeck}, {Schulte}, {Schutz}, {Schwartz},
  {Scott}, {Scott}, {Seglar-Arroyo}, {Seidel}, {Sekiguchi}, {Sekiguchi},
  {Sellers}, {Sengupta}, {Sennett}, {Sentenac}, {Seo}, {Sequino}, {Sergeev},
  {Setyawati}, {Shaffer}, {Shahriar}, {Shams}, {Shao}, {Sharifi}, {Sharma},
  {Sharma}, {Shawhan}, {Shcheblanov}, {Shen}, {Shibagaki}, {Shikauchi},
  {Shimizu}, {Shimoda}, {Shimode}, {Shink}, {Shinkai}, {Shishido}, {Shoda},
  {Shoemaker}, {Shoemaker}, {Shukla}, {Shyamsundar}, {Sieniawska}, {Sigg},
  {Singer}, {Singh}, {Singh}, {Singha}, {Sintes}, {Sipala}, {Skliris},
  {Slagmolen}, {Slaven-Blair}, {Smetana}, {Smith}, {Smith}, {Somala}, {Somiya},
  {Son}, {Soni}, {Soni}, {Sorazu}, {Sordini}, {Sorrentino}, {Sorrentino},
  {Sotani}, {Soulard}, {Souradeep}, {Sowell}, {Spagnuolo}, {Spencer}, {Spera},
  {Srivastava}, {Srivastava}, {Staats}, {Stachie}, {Steer}, {Steinlechner},
  {Steinlechner}, {Stops}, {Stevenson}, {Stover}, {Strain}, {Strang},
  {Stratta}, {Strunk}, {Sturani}, {Stuver}, {S{\"u}dbeck}, {Sudhagar},
  {Sudhir}, {Sugimoto}, {Suh}, {Summerscales}, {Sun}, {Sun}, {Sunil}, {Sur},
  {Suresh}, {Sutton}, {Suzuki}, {Suzuki}, {Swinkels}, {Szczepa{\'n}czyk},
  {Szewczyk}, {Tacca}, {Tagoshi}, {Tait}, {Takahashi}, {Takahashi}, {Takamori},
  {Takano}, {Takeda}, {Takeda}, {Talbot}, {Tanaka}, {Tanaka}, {Tanaka},
  {Tanaka}, {Tanaka}, {Tanasijczuk}, {Tanioka}, {Tanner}, {Tao}, {Tapia},
  {Tapia San Martin}, {Tasson}, {Telada}, {Tenorio}, {Terkowski}, {Test},
  {Thirugnanasambandam}, {Thomas}, {Thomas}, {Thompson}, {Thondapu}, {Thorne},
  {Thrane}, {Tiwari}, {Tiwari}, {Tiwari}, {Toland}, {Tolley}, {Tomaru},
  {Tomigami}, {Tomura}, {Tonelli}, {Torres-Forn{\'e}}, {Torrie}, {Tosta E
  Melo}, {T{\"o}yr{\"a}}, {Trapananti}, {Travasso}, {Traylor}, {Tringali},
  {Tripathee}, {Troiano}, {Trovato}, {Trozzo}, {Trudeau}, {Tsai}, {Tsai},
  {Tsang}, {Tsang}, {Tsao}, {Tse}, {Tso}, {Tsubono}, {Tsuchida}, {Tsukada},
  {Tsuna}, {Tsutsui}, {Tsuzuki}, {Turconi}, {Tuyenbayev}, {Ubhi}, {Uchikata},
  {Uchiyama}, {Udall}, {Ueda}, {Uehara}, {Ueno}, {Ueshima}, {Ugolini},
  {Unnikrishnan}, {Uraguchi}, {Urban}, {Ushiba}, {Usman}, {Utina}, {Vahlbruch},
  {Vajente}, {Vajpeyi}, {Valdes}, {Valentini}, {Valsan}, {van Bakel}, {van
  Beuzekom}, {van den Brand}, {van den Broeck}, {Vander-Hyde}, {van der
  Schaaf}, {van Heijningen}, {Vanosky}, {van Putten}, {Vardaro}, {Vargas},
  {Varma}, {Vas{\'u}th}, {Vecchio}, {Vedovato}, {Veitch}, {Veitch},
  {Venkateswara}, {Venneberg}, {Venugopalan}, {Verkindt}, {Verma}, {Veske},
  {Vetrano}, {Vicer{\'e}}, {Viets}, {Villa-Ortega}, {Vinet}, {Vitale}, {Vo},
  {Vocca}, {von Reis}, {von Wrangel}, {Vorvick}, {Vyatchanin}, {Wade}, {Wade},
  {Wagner}, {Walet}, {Walker}, {Wallace}, {Wallace}, {Walsh}, {Wang}, {Wang},
  {Wang}, {Ward}, {Warner}, {Was}, {Washimi}, {Washington}, {Watchi}, {Weaver},
  {Wei}, {Weinert}, {Weinstein}, {Weiss}, {Weller}, {Wellmann}, {Wen},
  {We{\ss}els}, {Westhouse}, {Wette}, {Whelan}, {White}, {Whiting}, {Whittle},
  {Wilken}, {Williams}, {Williams}, {Williamson}, {Willis}, {Willke}, {Wilson},
  {Winkler}, {Wipf}, {Wlodarczyk}, {Woan}, {Woehler}, {Wofford}, {Wong}, {Wu},
  {Wu}, {Wu}, {Wu}, {Wysocki}, {Xiao}, {Xu}, {Yamada}, {Yamamoto}, {Yamamoto},
  {Yamamoto}, {Yamamoto}, {Yamashita}, {Yamazaki}, {Yang}, {Yang}, {Yang},
  {Yang}, {Yang}, {Yap}, {Yeeles}, {Yelikar}, {Ying}, {Yokogawa}, {Yokoyama},
  {Yokozawa}, {Yoon}, {Yoshioka}, {Yu}, {Yu}, {Yuzurihara}, {Zadro{\.z}ny},
  {Zanolin}, {Zappa}, {Zeidler}, {Zelenova}, {Zendri}, {Zevin}, {Zhan},
  {Zhang}, {Zhang}, {Zhang}, {Zhang}, {Zhang}, {Zhao}, {Zhao}, {Zhao}, {Zhao},
  {Zhou}, {Zhu}, {Zhu}, {Zimmerman}, {Zlochower}, {Zucker}, {Zweizig}, {Ligo
  Scientific Collaboration}, {VIRGO Collaboration}, \& {KAGRA
  Collaboration}}]{LIGO_NSBHs:2021}
---. 2021, \apjl, 915, L5

\bibitem[{{Amaro-Seoane} \& et~al.(2017)}]{LISA:2017}
{Amaro-Seoane}, P., \& et~al. 2017, ArXiv e-prints, arXiv:1702.00786

\bibitem[{{Antoniadis} {et~al.}(2021){Antoniadis}, {Aguilera-Dena},
  {Vigna-G{\'o}mez}, {Kramer}, {Langer}, {M{\"u}ller}, {Tauris}, {Wang}, \&
  {Xu}}]{Antoniadis+2021}
{Antoniadis}, J., {Aguilera-Dena}, D.~R., {Vigna-G{\'o}mez}, A., {et~al.} 2021,
  arXiv e-prints, arXiv:2110.01393

\bibitem[{{Band} {et~al.}(1993){Band}, {Matteson}, {Ford}, {Schaefer},
  {Palmer}, {Teegarden}, {Cline}, {Briggs}, {Paciesas}, {Pendleton}, {Fishman},
  {Kouveliotou}, {Meegan}, {Wilson}, \& {Lestrade}}]{Band:1993}
{Band}, D., {Matteson}, J., {Ford}, L., {et~al.} 1993, \apj, 413, 281

\bibitem[{{Bildsten} \& {Cutler}(1992)}]{BildstenCutler:1992}
{Bildsten}, L., \& {Cutler}, C. 1992, \apj, 400, 175

\bibitem[{{Blackburn} {et~al.}(2015){Blackburn}, {Briggs}, {Camp},
  {Christensen}, {Connaughton}, {Jenke}, {Remillard}, \&
  {Veitch}}]{Blackburn+2015}
{Blackburn}, L., {Briggs}, M.~S., {Camp}, J., {et~al.} 2015, \apjs, 217, 8

\bibitem[{{Bransgrove} {et~al.}(2018){Bransgrove}, {Levin}, \&
  {Beloborodov}}]{BransgroveLevinBelo_2018}
{Bransgrove}, A., {Levin}, Y., \& {Beloborodov}, A. 2018, \mnras, 473, 2771

\bibitem[{{Carrasco} {et~al.}(2021){Carrasco}, {Shibata}, \&
  {Reula}}]{Carrasco+2021}
{Carrasco}, F., {Shibata}, M., \& {Reula}, O. 2021, \prd, 104, 063004

\bibitem[{{Chen} \& {Dai}(2021)}]{ChenDai:2021}
{Chen}, K., \& {Dai}, Z.~G. 2021, \apj, 909, 4

\bibitem[{{Dai}(2019)}]{Dai_NSBHQ:2019}
{Dai}, Z.~G. 2019, \apjl, 873, L13

\bibitem[{{D'Orazio} \& {Levin}(2013)}]{DL:2013}
{D'Orazio}, D.~J., \& {Levin}, J. 2013, \prd, 88, 064059

\bibitem[{{D'Orazio} {et~al.}(2016){D'Orazio}, {Levin}, {Murray}, \&
  {Price}}]{DL:2016}
{D'Orazio}, D.~J., {Levin}, J., {Murray}, N.~W., \& {Price}, L. 2016, \prd, 94,
  023001

\bibitem[{{Elfritz} {et~al.}(2016){Elfritz}, {Pons}, {Rea}, {Glampedakis}, \&
  {Vigan{\`o}}}]{Elfritz+2016}
{Elfritz}, J.~G., {Pons}, J.~A., {Rea}, N., {Glampedakis}, K., \& {Vigan{\`o}},
  D. 2016, \mnras, 456, 4461

\bibitem[{{Espinoza} {et~al.}(2011){Espinoza}, {Lyne}, {Kramer}, {Manchester},
  \& {Kaspi}}]{EspinozaLyne+2011}
{Espinoza}, C.~M., {Lyne}, A.~G., {Kramer}, M., {Manchester}, R.~N., \&
  {Kaspi}, V.~M. 2011, \apjl, 741, L13

\bibitem[{{Foucart} {et~al.}(2018){Foucart}, {Hinderer}, \&
  {Nissanke}}]{Foucart:2018}
{Foucart}, F., {Hinderer}, T., \& {Nissanke}, S. 2018, \prd, 98, 081501

\bibitem[{{Fragione}(2021)}]{FragioneNSBH:2021}
{Fragione}, G. 2021, arXiv e-prints, arXiv:2110.09604

\bibitem[{{GCN archive for LIGO/Virgo S190814bv}(2020)}]{GCN_190814}
{GCN archive for LIGO/Virgo S190814bv}. 2020,
  \url{https://gcn.gsfc.nasa.gov/other/GW190814bv.gcn3}, accessed: 2021-10-20

\bibitem[{{GCN archive for S200105ae}(2020)}]{GCN_105}
{GCN archive for S200105ae}. 2020,
  \url{https://gcn.gsfc.nasa.gov/other/S200105ae.gcn3}, accessed: 2021-09-21

\bibitem[{{GCN archive for S200115j}(2020)}]{GCN_115}
{GCN archive for S200115j}. 2020,
  \url{https://gcn.gsfc.nasa.gov/other/S200115j.gcn3}, accessed: 2021-09-21

\bibitem[{{Geppert} {et~al.}(1999){Geppert}, {Page}, \&
  {Zannias}}]{HiddenMagnetars:1999}
{Geppert}, U., {Page}, D., \& {Zannias}, T. 1999, \aap, 345, 847

\bibitem[{{Goldreich} \& {Lynden-Bell}(1969)}]{GLB:1969}
{Goldreich}, P., \& {Lynden-Bell}, D. 1969, \apj, 156, 59

\bibitem[{{Goldstein} {et~al.}(2016){Goldstein}, {Burns}, {Hamburg},
  {Connaughton}, {Veres}, {Briggs}, {Hui}, \& {The GBM-LIGO
  Collaboration}}]{O2GBM_searchupdate:2016}
{Goldstein}, A., {Burns}, E., {Hamburg}, R., {et~al.} 2016, arXiv e-prints,
  arXiv:1612.02395

\bibitem[{{Gusakov} {et~al.}(2020){Gusakov}, {Kantor}, \&
  {Ofengeim}}]{Gusakov+2020}
{Gusakov}, M.~E., {Kantor}, E.~M., \& {Ofengeim}, D.~D. 2020, \mnras, 499, 4561

\bibitem[{{Hansen} \& {Lyutikov}(2001)}]{HansenLyut:2001}
{Hansen}, B. M.~S., \& {Lyutikov}, M. 2001, \mnras, 322, 695

\bibitem[{{Igoshev} {et~al.}(2021){Igoshev}, {Hollerbach}, {Wood}, \&
  {Gourgouliatos}}]{Igoshev+2021}
{Igoshev}, A.~P., {Hollerbach}, R., {Wood}, T., \& {Gourgouliatos}, K.~N. 2021,
  Nature Astronomy, 5, 145

\bibitem[{{Isoyama} {et~al.}(2018){Isoyama}, {Nakano}, \&
  {Nakamura}}]{DECIGO_MB:2018}
{Isoyama}, S., {Nakano}, H., \& {Nakamura}, T. 2018, Progress of Theoretical
  and Experimental Physics, 2018, 073E01

\bibitem[{{Kaspi}(2010)}]{Kaspi:2010}
{Kaspi}, V.~M. 2010, Proceedings of the National Academy of Science, 107, 7147

\bibitem[{{Lai}(2012)}]{DLai:2012}
{Lai}, D. 2012, \apjl, 757, L3

\bibitem[{{Laine} \& {Lin}(2012)}]{LaineLin:2012}
{Laine}, R.~O., \& {Lin}, D. N.~C. 2012, \apj, 745, 2

\bibitem[{{Lee}(1993)}]{1993ApJ...418..147L}
{Lee}, M.~H. 1993, \apj, 418, 147

\bibitem[{{Levin} {et~al.}(2018){Levin}, {D'Orazio}, \&
  {Garcia-Saenz}}]{LDG:2018}
{Levin}, J., {D'Orazio}, D.~J., \& {Garcia-Saenz}, S. 2018, \prd, 98, 123002

\bibitem[{{Liu} {et~al.}(2020){Liu}, {Hu}, {Zhang}, \& {Mei}}]{TianQinMB:2020}
{Liu}, S., {Hu}, Y.-M., {Zhang}, J.-d., \& {Mei}, J. 2020, \prd, 101, 103027

\bibitem[{{Luo} {et~al.}(2016){Luo}, {Chen}, {Duan}, {Gong}, {Hu}, {Ji}, {Liu},
  {Mei}, {Milyukov}, {Sazhin}, {Shao}, {Toth}, {Tu}, {Wang}, {Wang}, {Yeh},
  {Zhan}, {Zhang}, {Zharov}, \& {Zhou}}]{TianQin:2016}
{Luo}, J., {Chen}, L.-S., {Duan}, H.-Z., {et~al.} 2016, Classical and Quantum
  Gravity, 33, 035010

\bibitem[{{Lyutikov}(2011)}]{Lyut:2011}
{Lyutikov}, M. 2011, \prd, 83, 064001

\bibitem[{{Lyutikov} \& {McKinney}(2011)}]{LyutikovMckinney:2011}
{Lyutikov}, M., \& {McKinney}, J.~C. 2011, \prd, 84, 084019

\bibitem[{{Mandel} \& {Broekgaarden}(2021)}]{2021arXiv210714239M}
{Mandel}, I., \& {Broekgaarden}, F.~S. 2021, arXiv e-prints, arXiv:2107.14239

\bibitem[{{Mandel} \& {Smith}(2021)}]{MandelSmith:2021}
{Mandel}, I., \& {Smith}, R. J.~E. 2021, \apjl, 922, L14

\bibitem[{{McKernan} {et~al.}(2020){McKernan}, {Ford}, \&
  {O'Shaughnessy}}]{McKernanNSBH_AGN:2020}
{McKernan}, B., {Ford}, K.~E.~S., \& {O'Shaughnessy}, R. 2020, \mnras, 498,
  4088

\bibitem[{{McWilliams} \& {Levin}(2011)}]{McL:2011}
{McWilliams}, S.~T., \& {Levin}, J. 2011, \apj, 742, 90

\bibitem[{{Mingarelli} {et~al.}(2017){Mingarelli}, {Lazio}, {Sesana}, {Greene},
  {Ellis}, {Ma}, {Croft}, {Burke-Spolaor}, \& {Taylor}}]{Mingarelli+2017}
{Mingarelli}, C.~M.~F., {Lazio}, T.~J.~W., {Sesana}, A., {et~al.} 2017, Nature
  Astronomy, 1, 886

\bibitem[{{Miralles} {et~al.}(2002){Miralles}, {Pons}, \&
  {Urpin}}]{Miralles+2002}
{Miralles}, J.~A., {Pons}, J.~A., \& {Urpin}, V.~A. 2002, \apj, 574, 356

\bibitem[{{Narayan} {et~al.}(1992){Narayan}, {Paczynski}, \&
  {Piran}}]{NarayanPacz:1992}
{Narayan}, R., {Paczynski}, B., \& {Piran}, T. 1992, \apjl, 395, L83

\bibitem[{{Neill} {et~al.}(2021){Neill}, {Tsang}, {van Eerten}, {Ryan}, \&
  {Newton}}]{DuncanTsang+2021}
{Neill}, D., {Tsang}, D., {van Eerten}, H., {Ryan}, G., \& {Newton}, W.~G.
  2021, arXiv e-prints, arXiv:2111.03686

\bibitem[{{Paczynski}(1986)}]{Pacz:1986GRB}
{Paczynski}, B. 1986, \apjl, 308, L43

\bibitem[{{Pan} \& {Yang}(2019)}]{PanYang_BHdischarge:2019}
{Pan}, Z., \& {Yang}, H. 2019, \prd, 100, 043025

\bibitem[{{Paschalidis} {et~al.}(2013){Paschalidis}, {Etienne}, \&
  {Shapiro}}]{Paschalidis+2013}
{Paschalidis}, V., {Etienne}, Z.~B., \& {Shapiro}, S.~L. 2013, \prd, 88, 021504

\bibitem[{{Piro}(2012)}]{Piro:2012}
{Piro}, A.~L. 2012, \apj, 755, 80

\bibitem[{{Sesana}(2016)}]{Sesana:LISALIGO:2016}
{Sesana}, A. 2016, \prl, 116, 231102

\bibitem[{{Stevenson} {et~al.}(2017){Stevenson}, {Vigna-G{\'o}mez}, {Mandel},
  {Barrett}, {Neijssel}, {Perkins}, \& {de Mink}}]{2017NatCo...814906S}
{Stevenson}, S., {Vigna-G{\'o}mez}, A., {Mandel}, I., {et~al.} 2017, Nature
  Communications, 8, 14906

\bibitem[{{Tagawa} {et~al.}(2021){Tagawa}, {Kocsis}, {Haiman}, {Bartos},
  {Omukai}, \& {Samsing}}]{TagawaKocsis+2021}
{Tagawa}, H., {Kocsis}, B., {Haiman}, Z., {et~al.} 2021, \apj, 908, 194

\bibitem[{{Team COMPAS} {et~al.}(2021){Team COMPAS}, {Riley}, {Agrawal},
  {Barrett}, {Boyett}, {Broekgaarden}, {Chattopadhyay}, {Gaebel}, {Gittins},
  {Hirai}, {Howitt}, {Justham}, {Khandelwal}, {Kummer}, {Lau}, {Mandel}, {de
  Mink}, {Neijssel}, {Riley}, {van Son}, {Stevenson}, {Vigna-Gomez},
  {Vinciguerra}, {Wagg}, \& {Willcox}}]{2021arXiv210910352T}
{Team COMPAS}, {Riley}, J., {Agrawal}, P., {et~al.} 2021, arXiv e-prints,
  arXiv:2109.10352

\bibitem[{{Thorne} {et~al.}(1986){Thorne}, {Price}, \& {MacDonald}}]{MPBook}
{Thorne}, K.~S., {Price}, R.~H., \& {MacDonald}, D.~A. 1986, {Black holes: The
  membrane paradigm} (Yale University Press)

\bibitem[{{Tiengo} {et~al.}(2013){Tiengo}, {Esposito}, {Mereghetti}, {Turolla},
  {Nobili}, {Gastaldello}, {G{\"o}tz}, {Israel}, {Rea}, {Stella}, {Zane}, \&
  {Bignami}}]{Tiengo+2013}
{Tiengo}, A., {Esposito}, P., {Mereghetti}, S., {et~al.} 2013, \nat, 500, 312

\bibitem[{{Vietri}(1996)}]{Vietri1996}
{Vietri}, M. 1996, \apjl, 471, L95

\bibitem[{{Vigna-G{\'o}mez} {et~al.}(2018){Vigna-G{\'o}mez}, {Neijssel},
  {Stevenson}, {Barrett}, {Belczynski}, {Justham}, {de Mink}, {M{\"u}ller},
  {Podsiadlowski}, {Renzo}, {Sz{\'e}csi}, \& {Mandel}}]{2018MNRAS.481.4009V}
{Vigna-G{\'o}mez}, A., {Neijssel}, C.~J., {Stevenson}, S., {et~al.} 2018,
  \mnras, 481, 4009

\bibitem[{{Vigna-G{\'o}mez} {et~al.}(2020){Vigna-G{\'o}mez}, {MacLeod},
  {Neijssel}, {Broekgaarden}, {Justham}, {Howitt}, {de Mink}, {Vinciguerra}, \&
  {Mandel}}]{2020PASA...37...38V}
{Vigna-G{\'o}mez}, A., {MacLeod}, M., {Neijssel}, C.~J., {et~al.} 2020, \pasa,
  37, e038

\bibitem[{Vigna-Gómez {et~al.}(2021)Vigna-Gómez, Aguilera-Dena, \&
  Willcox}]{vigna_gomez_alejandro_2021_4682798}
Vigna-Gómez, A., Aguilera-Dena, D.~R., \& Willcox, R. 2021, {Dataset from:
  Fallback Supernova Assembly of Heavy Binary Neutron Stars and Light Black
  Hole-Neutron Star Pairs and the Common Stellar Ancestry of GW190425 and
  GW200115}, doi:10.5281/zenodo.4682798

\bibitem[{{Wald}(1974)}]{Wald:1974}
{Wald}, R.~M. 1974, \prd, 10, 1680

\bibitem[{{Xu} {et~al.}(2021){Xu}, {Li}, {Cui}, {Li}, {Shao}, {Liang}, \&
  {Liu}}]{MagnetarsinXRBs:2021}
{Xu}, K., {Li}, X.-D., {Cui}, Z., {et~al.} 2021, arXiv e-prints,
  arXiv:2110.10438

\bibitem[{{Yang} {et~al.}(2020{\natexlab{a}}){Yang}, {Gayathri}, {Bartos},
  {Haiman}, {Safarzadeh}, \& {Tagawa}}]{Yang_Gayathri+2020}
{Yang}, Y., {Gayathri}, V., {Bartos}, I., {et~al.} 2020{\natexlab{a}}, \apjl,
  901, L34

\bibitem[{{Yang} {et~al.}(2020{\natexlab{b}}){Yang}, {Zhong}, {Zhang}, {Wu},
  {Zhang}, {Yang}, {Cao}, {Gao}, {Zou}, {Wang}, {L{\"u}}, {Cang}, \&
  {Dai}}]{Yang+2020}
{Yang}, Y.-S., {Zhong}, S.-Q., {Zhang}, B.-B., {et~al.} 2020{\natexlab{b}},
  \apj, 899, 60

\bibitem[{{Ye} {et~al.}(2020){Ye}, {Fong}, {Kremer}, {Rodriguez}, {Chatterjee},
  {Fragione}, \& {Rasio}}]{Ye_GCNSBH+2020}
{Ye}, C.~S., {Fong}, W.-f., {Kremer}, K., {et~al.} 2020, \apjl, 888, L10

\bibitem[{{Zhang}(2019)}]{BZang_ChargedEMC:2019}
{Zhang}, B. 2019, \apjl, 873, L9

\bibitem[{{Zhong} {et~al.}(2019){Zhong}, {Dai}, \& {Deng}}]{ZhongDD+2019}
{Zhong}, S.-Q., {Dai}, Z.-G., \& {Deng}, C.-M. 2019, \apjl, 883, L19

\bibitem[{{Zrake} \& {MacFadyen}(2013)}]{ZrakeMacFad:2013}
{Zrake}, J., \& {MacFadyen}, A.~I. 2013, \apjl, 769, L29

\end{thebibliography}
\end{document}